# Strengthening in multi-principal element alloys with local-chemical-order roughened dislocation pathways

Qing-Jie Li[1], Howard Sheng[2,3*], Evan Ma[1*]

[1]Department of Materials Science and Engineering, Johns Hopkins University, Baltimore, MD 21218, USA;

[2]Department of Physics and Astronomy, George Mason University, Fairfax, Virginia 22030, USA.

[3]Center for High Pressure Science and Technology Advanced Research, Shanghai 201203, China

[*]To whom correspondence may be addressed.

Email: hsheng@gmu.edu; ema@jhu.edu

High-entropy alloys (HEAs) were presumed to have a configurational entropy as high as that of an ideally mixed solid solution (SS) of multiple elements in near-equal proportions. However, enthalpic interactions inevitably render such chemically disordered SSs rare and metastable, except at very high temperatures. Here we highlight a structural feature that sets these concentrated SSs apart from traditional solvent-solute ones: the HEAs possess a wide variety of *(local) chemical ordering* (LCO). Our atomistic simulations employing an empirical interatomic potential for NiCoCr reveal that the LCO of the multi-principal-element SS changes conspicuously with alloy processing conditions, producing a wide range of generalized planar fault energy in terms of both its sample-average and spatial variation. We further demonstrate that the LCO heightens the ruggedness of the energy landscape and raises activation barriers governing dislocation activities. This not only influences the selection of dislocation pathways in slip, faulting, twinning, and martensitic transformation, but also increases the lattice friction to dislocation motion via a new mechanism of nanoscale segment detrapping that elevates the mechanical strength. All these open a vast playground not accessible to ground-state SSs or intermetallics, offering rich opportunities to tune properties.

Multi-principal element materials, with compositions in the central region of the multicomponent phase diagram, are currently dubbed as "high-entropy" alloys (HEAs). These complex concentrated alloys are emerging as a new research field attracting considerable attention.[1–11] However, a fundamental materials science question remains outstanding in the community, namely, what is the new tell-tale feature that distinguishes these HEAs from traditional solid solution (SS) alloys. The original answer was that the HEAs correspond to an unusually high configurational entropy of mixing ($S_c$)[1], with a magnitude >~1.61$R$ estimated from $S_{c,ideal} = -R\sum_{i=1}^{N} x_i \ln(x_i)$ for ideal solutions, where $R$ is the gas constant and $x_i$ is the molar fraction of the *ith* component. As of today, such a random solid solution (RSS) picture remains a common assumption in the HEA community. However, this RSS state is possible only at very high temperatures, where the degree of local chemical order (LCO) is negligible and the entropy term predominates the free energy reduction to dictate ideal mixing. Real-world HEAs are actually processed (annealed and homogenized) and used at relatively low temperatures,[10,12–18] where complex enthalpic interactions among various constituent elements come into play. Even the pioneering Cantor alloy, a face-centered-cubic (FCC) SS close to random if annealed at 1100 $^{o}$C,[19] decomposes after long anneal at < 900 $^{o}$C.[17] Appreciable chemical order has in fact been found in a number of HEAs.[20–27] As such, HEAs, even when in the form of a single-phase SS, can be of a mixing entropy far below $S_{c,ideal}$.

In this paper we advocate a different perspective on HEAs: they are special in the "vastness of possible LCO configurations", and as such even a given HEA composition can deliver a plethora of properties. Specifically, a concentrated SS ushers in an unprecedentedly large variability of LCOs, from the extreme of (overly simplified) RSS, all the way to fully ordered ground-state intermetallics. These intermediate states highlight HEA's metastable nature, and are beyond reach in traditional (terminal) SSs, which can instead be approximated as Raoult's or Henry's solution, where solutes always sparsely and randomly distribute in a solvent crystal lattice, producing only one specific set of properties. Now for concentrated HEAs, the type (species involved), degree (magnitude of the order parameter) and extent (length scale and spatial distribution) of LCO all span a wide range. In what follows we will systematically demonstrate, using an atomistic model that mimics the face-centered-cubic (FCC) NiCoCr medium-entropy alloy (MEA), the new features rendered by the concentrated compositions: complex stacking fault energy that is varying spatially and with processing conditions, local antiphase boundary energy, and unconventional twin boundary energy and martensite phase boundary energy.

This rugged energy landscape heightens the lattice resistance to dislocation motion. We will show that the presence of multi-principal elements and their variable LCO make dislocations behave differently from conventional metals and dilute solutions, in terms of slip paths and activated nanoscale segment detrapping processes that govern alloy strength. The multitude of (partial) dislocation behavior associated with the adjustable LCOs opens new opportunities to tailor properties, through judicious choice of processing parameters, in particular, the homogenization annealing temperature ($T_a$). Our analysis provides an atomistic explanation for the experimentally observed strength elevation of HEAs, e.g., the lattice friction stress of NiCoCr was found several times higher than that of elemental FCC metals[28]; and the yield strength of TaNbHfZr was increased by up to 76% via annealing induced ordering at various intermediate temperatures[29].

**Variable LCOs in samples processed at different temperatures**

We first demonstrate the vast range of LCO in a concentrated SS of a given composition. We chose the FCC MEA NiCoCr as a model because it is a representative of multi-principal element systems and its mechanical properties are typical of (and often better than) other quaternary and quinary HEAs.[16,30] To enable insightful large-scale molecular dynamics (MD) simulations, which are sorely needed to provide atomistic insight about dislocation behavior and adequate statistics but have thus far been lacking in the HEA field, we developed an empirical interatomic potential for nonmagnetic NiCoCr (Methods). This model is designed to capture the typical features of HEAs: multi-principal constituents (equi-atomic composition) with similar atomic sizes, chemical interactions consistent with typical MEA solutions, single-phase solution but with variable LCO. A comprehensive discussion of this new potential is presented in Fig. S1-S7, Table S1-S2, as well as Section 12 in Supporting Information. We then carried out hybrid MD and Monte Carlo (MC) simulations (Methods) to obtain equilibrium configurations after annealing at different temperatures, $T_a$.

Fig. 1 shows the LCO for samples annealed at different $T_a$, with LCO measured by the pairwise multicomponent short-range order parameter[31] (Methods). As seen in Fig. 1a, with decreasing $T_a$, the absolute values of $\alpha^1_{\text{Ni-Ni}}$, $\alpha^1_{\text{Ni-Co}}$, $\alpha^1_{\text{Ni-Cr}}$, and $\alpha^1_{\text{Co-Cr}}$ first smoothly increase when $T_a$ > ~850 K and then dramatically rise to saturated values at sufficiently low $T_a$. Meanwhile, $\alpha^1_{\text{Co-Co}}$ and $\alpha^1_{\text{Cr-Cr}}$ are slightly negative and largely remain constant for the whole $T_a$ range. This suggests that our model NiCoCr system develops local Ni segregation and Co-Cr ordering with decreasing $T_a$ (see Fig. S8 for $\alpha^2$ and $\alpha^3$). This tendency is consistent with the equilibrium phase diagram to form Co-Cr intermetallic phase(s)[32]. Such increasing LCO also implies significant deviations from the configurational entropy of an ideal solution $S_{c,\text{ideal}}$. Fig. 1b shows the dependence of $S_c$ on processing temperature $T_a$, based on the cluster variation method (CVM) with pair approximation[33]. The overall trend is similar to what Gao *et al*. reported using a similar method for other HEAs[34]. As seen, an MEA/HEA rarely reaches $S_{c,\text{ideal}}$, ~95% at best at the highest $T_a$ (1650 K). With decreasing $T_a$, $S_c$ turns away from $S_{c,\text{ideal}}$ fairly early and loses half of its magnitude when LCO becomes obvious (compare with Fig. 1a). As such, a truly random SS is only an extreme state of MEA/HEA and difficult to reach in practice. More commonly, an HEA at a given composition possesses *partial chemical order*.

Fig.1c-e show three representative atomic configurations. As seen, samples prepared at relatively high $T_a$, e.g., 1350 K (Fig. 1c) and 950 K (Fig. 1d), show nanoscale Ni clusters and interconnected Co-Cr clusters with relatively random compositions and orientations. Below $T_a$ ~ 850 K, e.g., at $T_a$ = 650 K (Fig. 1e), dramatic chemical ordering creates compositionally identical but orientationally distinguishable Co-Cr domains, as marked by the domain boundaries in Fig. 1e (dashed lines). Randomly distributed Ni nanoscale precipitates break up these Co-Cr domains into finer regions. This visually obvious LCO persists across the $T_a$ range examined here up to temperatures (e.g., $T_a$ = 1650 K) near the melting point. For relatively high annealing temperatures such as those in Fig.1c-d, we expect that the kinetics needed for chemical ordering would be accessible in typical laboratory experiments: here we emphasize that local ordering/clustering or even compositional decomposition has been recently reported in several HEAs[17,29,35–37]. However, when $T_a$ is too low for adequate aging the predicted chemical ordering may require a timescale much longer than the typical homogenization duration in experiments. Nevertheless, kinetically permitted, all the HEAs evolve toward the ground state (even the pioneering Cantor HEA[17] is no exception). So the partially ordered system is actually the norm for single-phase HEA solutions.

In this context, Fig. 1d represents a microstructure at the other end opposite to the RSS. Between the two, there is ample room for structural engineering.

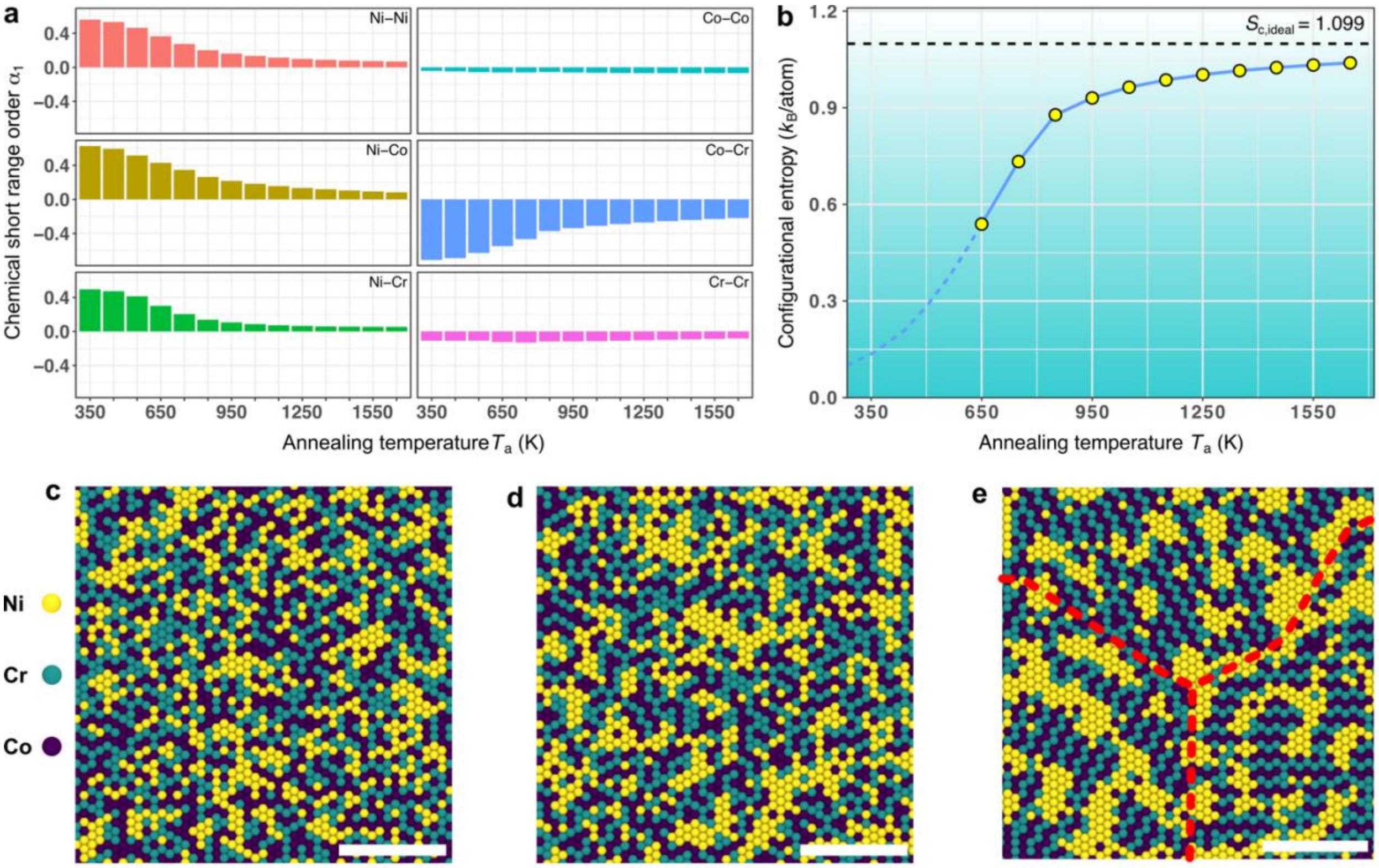

**Fig. 1** | Local chemical ordering at different annealing temperatures $T_a$. **a**, Pairwise chemical short-range order parameter $\alpha_1$ (Methods) at different annealing temperatures. **b**, Configurational entropy of the NiCoCr ternary solution and its temperature dependence. Points are data estimated through the cluster variation method (CVM) with pair approximation, connected using a blue line as guide for the eye. This approximation becomes increasingly inadequate at high LCO; thus a dashed line is used instead to project the trend at low $T_a$. Black dashed line denotes the $S_{c,ideal}$. **c-e**, Representative configurations at $T_a$ = 1350 K, $T_a$ = 950 K and $T_a$ = 650 K, respectively. The red dashed lines indicate the Co-Cr domain boundaries. The scale bar is 3 nm. All atomic configurations are viewed on the (111) plane.

## Energy pathways of slip, twinning and martensitic transformation

We next illustrate how the dislocation behavior would change, in NiCoCr MEA samples with different LCOs. Fig. 2 shows the energy landscape calculated at zero Kelvin (see Methods) for the pathways of slip, deformation twinning (DT) and FCC → HCP (hexagonal close-packed) martensitic transformation (MT) to demonstrate the large impact of the LCOs. Fig. 2a shows the energy landscape of ordinary dislocation slip in samples prepared at different $T_a$. A full dislocation slip consists of a leading partial dislocation slip (the first humps in Fig. 2a) and a trailing partial dislocation slip (the second humps in Fig. 2a). In our case, the leading partial dislocation (B → δ along the $[2\bar{1}\bar{1}]$ direction) induces both structural (FCC → HCP) and chemical changes (LCO breaking), resulting in a convoluted stacking fault energy (SFE), i.e., a significant fraction of SFE is due to breaking LCOs. Such chemical contribution to SFE is expected to be ubiquitous in real-world HEAs. We therefore adopt the notation of complex stacking fault (CSF) for HEAs, similar to that in superalloys. The sample-averaged CSF energy (CSFE) varies remarkably depending on $T_a$. For example, the RSS extreme shows a negative CSFE of -24.0 mJ/m$^2$, in accord with previous density functional theory calculations[38–41]. The unstable stacking fault energy is also close to that reported previously[39,41,42]. However, once LCO kicks in (via annealing), the CSFE jumps to a positive range from ~1.4 mJ/m$^2$ to ~57 mJ/m$^2$, as $T_a$ varies from 1650 K (near the melting point) to 650 K. In other words, an FCC HEA/MEA can take a CSFE value out of a wide range[43] and does not necessarily have the low SFE anticipated[38–42,44]. In laboratory experiments, the measured SFE for NiCoCr [45] is ~20 mJ/m$^2$; however, a one-to-one comparison between our model calculations and experimental results is not advisable because a) the LCO information is unavailable for the experimental samples; b) experiments use different method in evaluating SFE, and c) our model uses an empirical potential. The trailing partial dislocation (δ → A along the $[1\bar{2}1]$ direction) recovers the structural change (HCP → FCC) and eliminates the CSF. However, it interrupts LCO (except RSS) and creates local antiphase boundaries (APB), as indicated by the non-zero fault energies (A, Fig. 2a). Specifically, the local APB energies (APBEs) increases from ~50 mJ/m$^2$ to ~112 mJ/m$^2$ when $T_a$ varies from 1650 K to 650 K. Such variable APBEs are expected to be common in HEAs/MEAs with noticeable LCOs[46,20,26,47,17,18,35], and play an important role in mechanical properties (more discussions later).

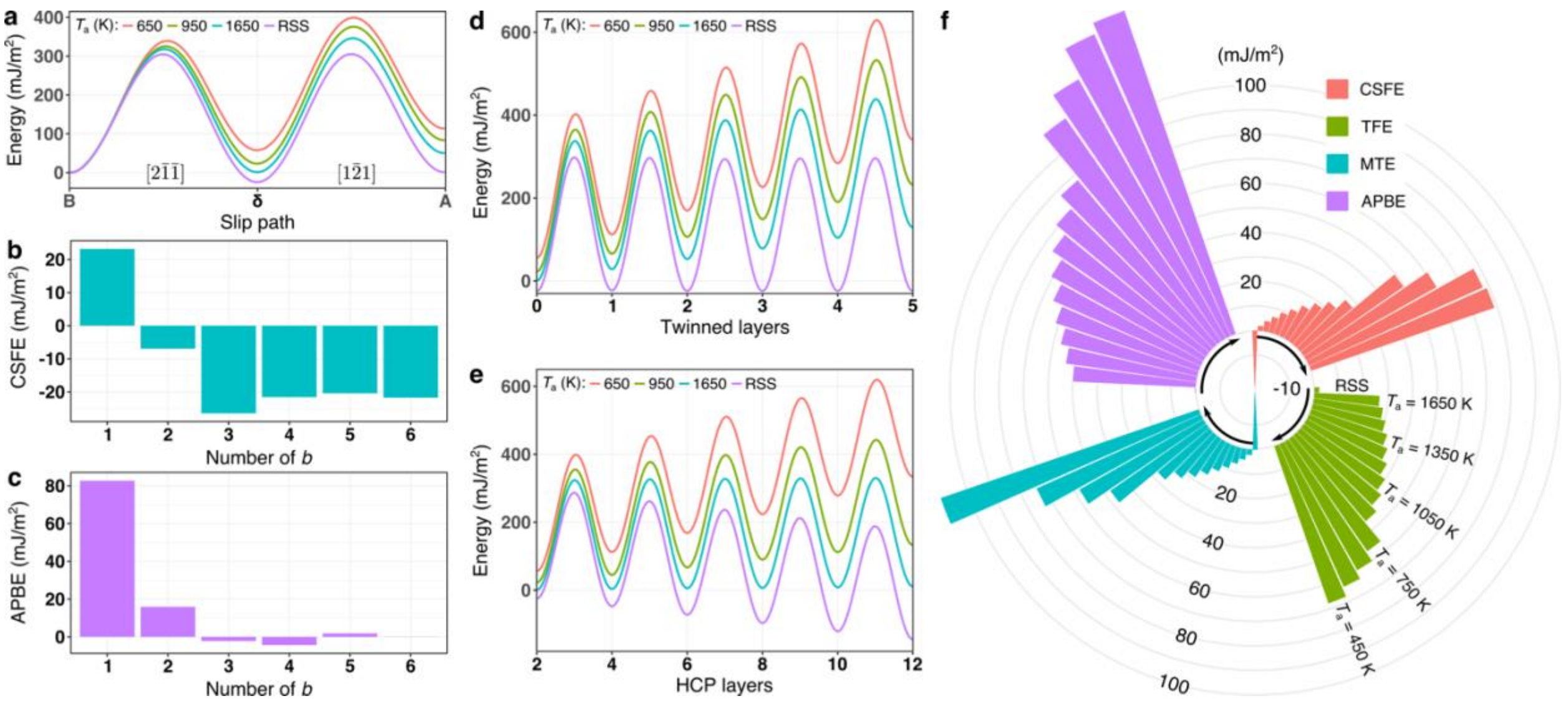

**Fig. 2 |** Energy landscape depends on the annealing temperature used to prepare the MEA. **a**, Full dislocation energy pathways for the first Burgers vector. **b-c**, CSFE and APBE changes upon increasing number of Burgers vector (on the same slip plane) for a sample with $T_a$ = 950 K. **d-e**, Energy pathways for twinning and FCC→HCP martensitic transformation, respectively. Both DT and MT are based on the CSF in **a**. **f**, The average complex stacking fault energy, twin fault energy, martensitic transformation energy and anti-phase boundary energy for RSS and samples with different $T_a$ (450 K – 1650K). The unit of the iso-energy circles is mJ/m$^2$.

Due to the short-to-medium range nature of LCO, both CSFE and APBE decay with repeated slip on the same slip plane. As shown in Fig. 2b and Fig. 2c, beyond 3*b* slip, the average CSFE and APBE decay to values similar to the RSS. This suggests that shear larger than 3*b* can destroy most of the LCO on the neighboring planes, leading to a slip plane softening mechanism[48] responsible for the experimentally observed planar slip. Plastic deformation is then a practical route to convert LCO to quasi-random state. See Fig. S9 for similar results on other samples.

DT and MT are also important routes of plastic deformation. Fig. 2d and Fig. 2e demonstrate the energy pathways of DT and the FCC → HCP MT. Twinning dislocations glide on consecutive slip planes while MT dislocations glide on every other plane, leading to growth steps of 1-layer and 2-layer, respectively. Apparently, the energy pathways of both DT and MT are functions of $T_a$ and the number of transformed layers $\lambda$ (the non-zero integer in Fig. 2d and Fig. 2e), $\gamma(T_a, \lambda)$. The boundary-energy (twin boundary or phase boundary) can then be defined as $\gamma_{\mathrm{B}} = \gamma(T_{\mathrm{a}}, \lambda)/2$. Again, at a given $\lambda$, samples with various LCOs produce a broad spectrum of $\gamma_{\mathrm{B}}$ for both DT and MT, with RSS serving as an extreme case. For DT in RSS, both the $\gamma_{\mathrm{B}}$ (~11 mJ/$m^2$) and unstable twin fault energy (~320 mJ/$m^2$) are similar to previous studies[39,41,42]. For MT in RSS, HCP phase growth significantly lowers $\gamma_{\mathrm{B}}$, suggesting the metastable nature of the FCC RSS. However, samples with LCOs show remarkably different behaviors. For example, $\gamma_{\mathrm{B}}$ in DT never converges to a constant value as $\lambda$ increases ($\gamma_{\mathrm{B}}$ increases approximately in a linear fashion with respect to $\lambda$, see Fig. S10), which is in sharp contrast to elemental FCC metals where $\gamma_{\mathrm{B}}$ generally converges to a constant value after several layers[49] such that the ensuing widening does not cost extra energy. Here for HEAs the unconventional twin boundary energies entail increasing energy penalty to break LCOs (associated with neighboring slip planes) as twin thickens. As a result, twin growth is no longer easy in HEAs, as the thickening always incurs additional energy penalty upon destroying the LCO layer by layer. Similar trend is also observed for the FCC → HCP MT. This offers an explanation to the experimental observation that separated SFs, nanotwins and very thin HCP laths dominate in deformed NiCoCr[30,42,45,50,51].

Fig. 2f summarizes and compares the average values (see Methods) of CSFE, twin fault energy (TFE, i.e., $\gamma(\lambda = 1) - \gamma(\lambda = 0)$ ), martensitic transformation energy (MTE, i.e., $\gamma(\lambda = 4) - \gamma(\lambda = 2)$) and APBE for samples with various LCO. Here again, the RSS serves as an extreme case characterized by negative CSFE, negative MTE, and negligible TFE and APBE. For samples annealed at $T_{\mathrm{a}} \geq 850$ K, the CSFE, TFE, MTE and APBE all increase with decreasing $T_{\mathrm{a}}$; however, their magnitude relative to one another remains similar to that of RSS, indicating a similarly strong tendency to form CSFs, nanotwins and thin HCP lamellae, consistent with the experimental observations[42,51]. For samples annealed below 850 K, all fault energies become relatively high and the energy costs for the leading and trailing partial dislocations become comparable, and thus narrowly extended dislocations may dominate the plastic deformation. See Fig. S11 for other LCO-dependent (or $T_{\mathrm{a}}$ dependent) material properties.

## Nanoscale heterogeneities due to spatial variations of LCO

The next important observation is that, in addition to the $T_{\mathrm{a}}$-dependent average LCO of a sample, inside a specific sample there is a spatial variation of LCOs creating various nanoscale heterogeneities, leading also to wide property distributions. In other words, the dislocation behavior is spatially variable on different length scales. In Fig. 3, we plot the local CSFEs and local APBEs to show the statistics and spatial variations (Methods).  Fig. 3a shows the probability density distributions of local CSFEs. As seen, for each processing condition, the local CSFEs exhibit significant spatial variations; the 25% – 75% accumulative probability range (highlighted by white boundaries) is generally greater than ~25 $mJ/m^2$, not to mention the even wider range between the minimum and maximum values. The distributions for relatively high $T_a$ ($\geq 950$ K) follow a Gaussian profile while low $T_a$ (e.g., 650 K) creates asymmetric distributions. The latter is closely related to the appreciable Co-Cr domains formed at relatively low $T_a$, i.e., each domain may have its own specific distribution (see Fig. S12 for an example) such that the merged overall distribution may no longer be Gaussian. In addition, the distributions shift to higher CSFE values with increasing LCO, consistent with the trend on the sample average in Fig. 2. Fig. 3b shows an example of spatially varying CSFEs in a sample with $T_a$ = 950 K. As seen, the CSFE is highly heterogeneous over space, with many nanoscale domains showing much lower/higher values than the average. As shown in Fig. 3c and Fig. 3d, due to the highly localized LCOs significant spatial variations are also observed for local APBEs, which, in contrast to intermetallic compounds such as the γ' phase in nickel superalloy, often exhibit significant deviations from the average value.

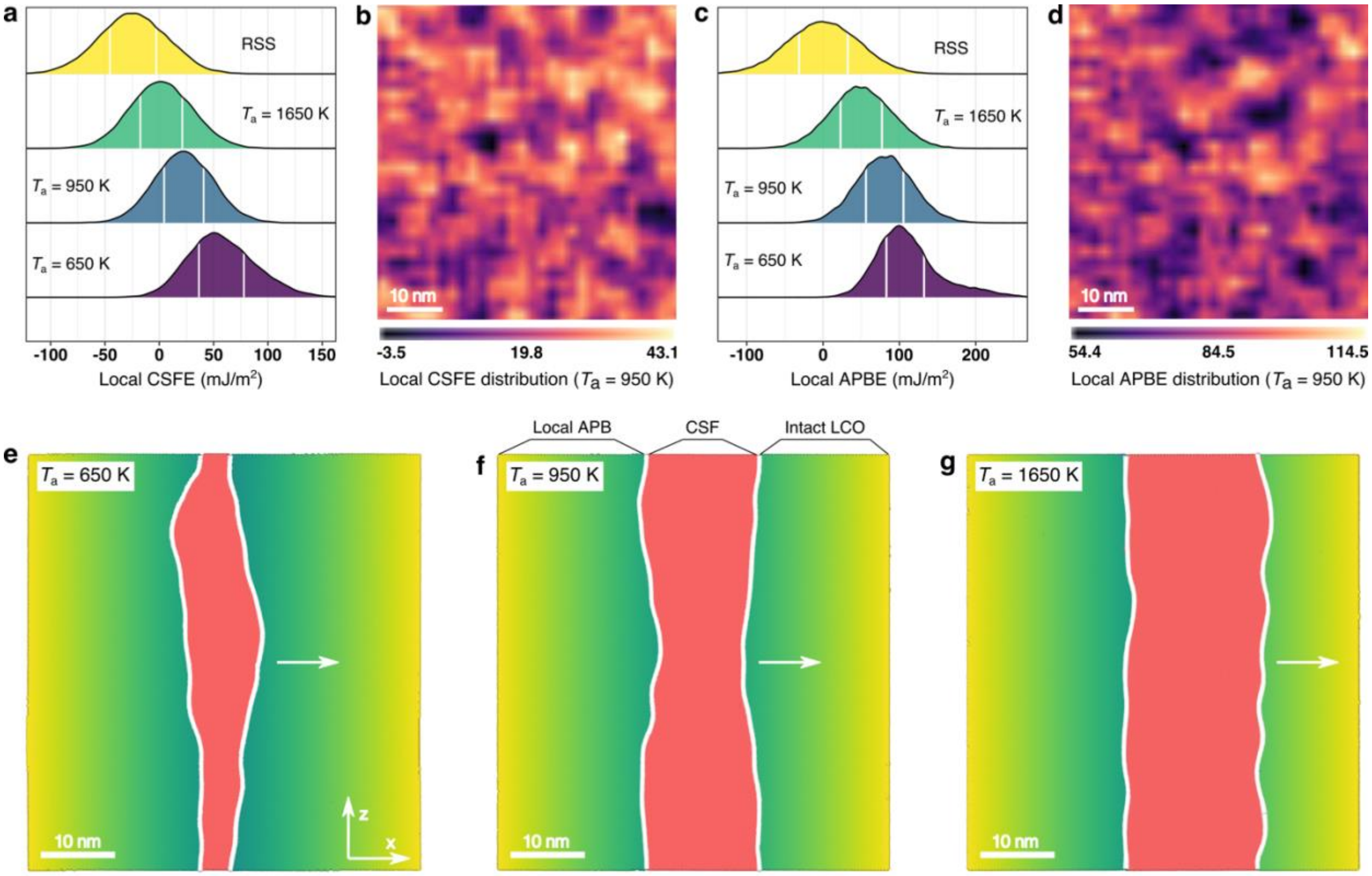

**Fig. 3 |** Statistical distributions of local properties and their effects on dislocation structure. **a**, Probability density distributions of local complex stacking fault energies for samples with $T_a$ = 650 K, 950 K, 1650 K and random solid solution. The 25% – 75% accumulative probability range is highlighted by white boundaries. **b**, Spatial distribution of local complex stacking fault energies in a sample with $T_a$ = 950 K. **c**, Probability density distributions of local antiphase boundary energies for samples with $T_a$ = 650 K, 950 K, 1650 K and random solid solution. **d**, Spatial distribution of local antiphase boundary energies in a sample with $T_a$ = 950 K. **e-g**, Examples of extended screw dislocation in samples with $T_a$ = 650 K, 950 K and 1650 K, respectively. All configurations in **e**, **f** and **g** are obtained after a 100 ps relaxation at $T$ = 300 K and under a constant shear stress of 300 MPa. The $x$, $z$ directions are along $[11\bar{2}]$ and $[1\bar{1}0]$, respectively. Periodic boundary conditions are only applied in $z$ direction. Dislocation cores are represented by the white tubes and CSFs are colored in red. Leading partial dislocation is on the right and trailing partial dislocation is on the left.

The chemical-order heterogeneities on the nanoscale are expected to present obstacles to dislocation movement and can be exploited to enhance the strength of HEAs. In Fig. 3e-g, we show three typical dislocation configurations in samples with $T_a$ = 650 K, 950 K and 1650 K, respectively. These dislocation configurations are the snapshots after 100 ps relaxation at 300 K and under a constant shear stress of 300 MPa (to counteract the restoring force due to the APBE). First of all, all samples show extended dislocations with apparently different average dissociation width that increases with increasing $T_a$, consistent with the trend of CSFE shown in Fig. 2. As a result, the dissociated dislocation configuration divides the crystal into three distinct regions (Fig. 3f), i.e., the local APB region due to full slip, the CSF region due to partial dislocation slip, and the intact LCO region without slip. Second, all partial dislocations exhibit nanoscale curvatures that vary along the dislocation lines, leading to wavy dislocation lines and rugged stacking fault ribbons. These results are consistent with the spatial variations of both CSFE and APBE shown in Fig. 3a-d, i.e., the dislocation line tends to bow out in 'soft' regions while 'hard' regions act as obstacles that trap dislocation segments. These spatial variations of dislocation configurations clearly demonstrate that the local properties span a wide range and some extreme values may be highly relevant to the rate-limiting step of a thermal activation process[52]. LCO is the root cause of the "local properties": the pronounced variations in dislocation splitting width and core structure have been reported in experimental samples[53,54] and in simulated RSS[55]. In this regard the HEA SS is more like a cocktail of many coexisting solid solutions.

**Dislocation motion and LCO-induced strengthening**

Now let us take a further step to examine the motion of dislocations that carry plastic flow, and demonstrate that LCOs indeed influence how difficult it is for dislocations to move and thus the strength of the alloy. To this end, we first resolve how a dislocation actually moves in a lattice with various LCOs. Here we consider a relatively long segment of a leading partial dislocation that often controls the mobility of an extended dislocation as well as the DT or MT processes. Fig. 4a shows a series of snapshots of the evolving dislocation line while it glides in the lattice, based on the MD simulation for Fig. 3f. Specifically, the leading partial dislocation is subject to a constant shear stress of 300 MPa at 300 K. As seen, the dislocation line is wavy and does not move smoothly, due to the nanoscale LCO heterogeneities as shown in Fig. 3; instead, the dislocation moves via a series of forward slip of local segments, each is on nanoscale and detraps from its local LCO environment. Highlighted in red in Fig. 4a are the nanoscale swept areas between the start and final states (e.g., see snapshot at 20 ps and 60 ps, respectively) corresponding to a particular nanoscale segment detrapping (NSD) event. This NSD, one at a time in an intermittent manner and one next to another locally on the dislocation line, is in sharp contrast to conventional FCC metals where dislocations move smoothly by either simultaneously propagating a long dislocation line (see Fig. S13 in Supporting Information for an example of Cu), or bow out in between unshearable obstacles that are spaced quite some distance (at least many nanometers) apart.

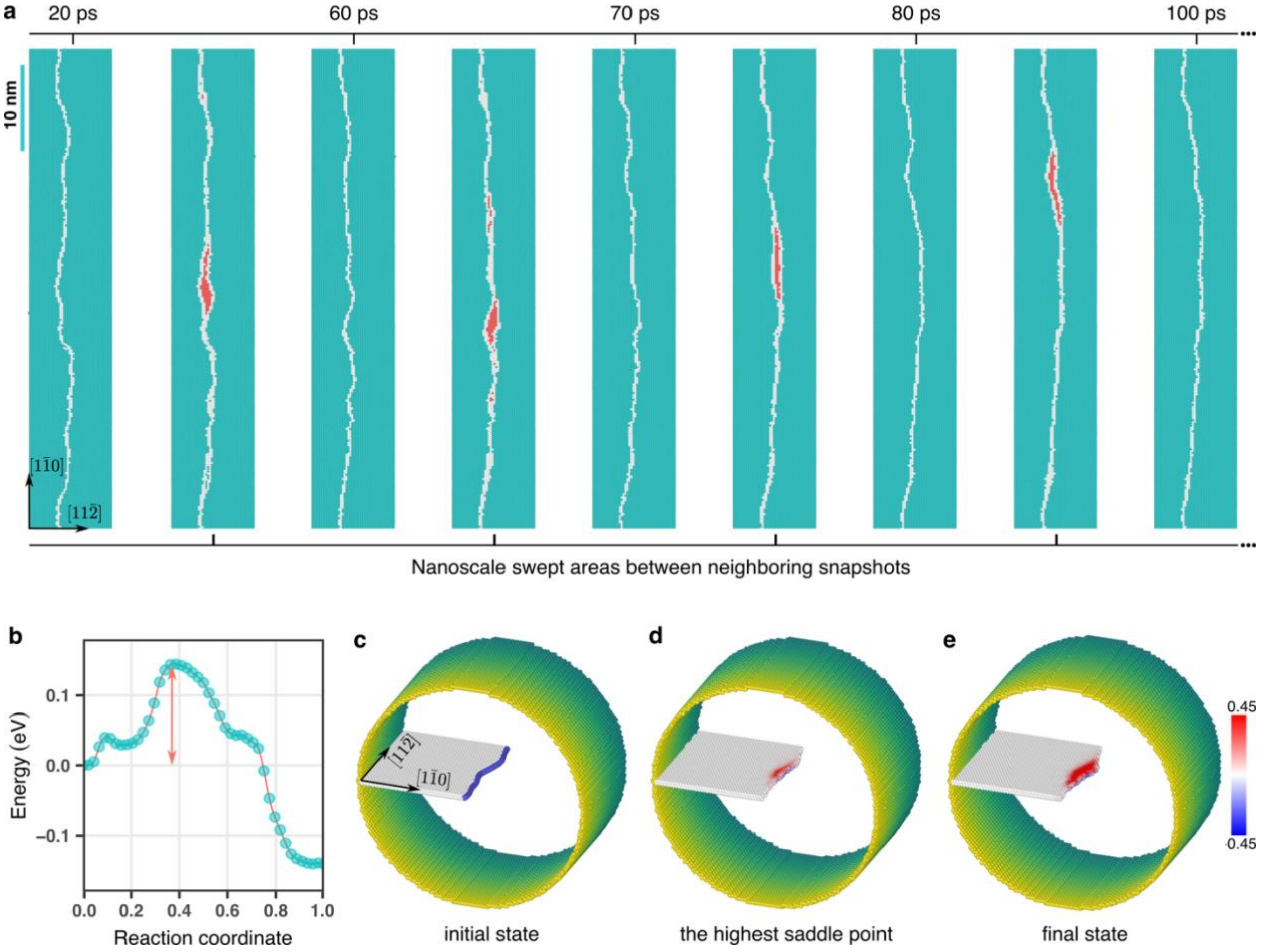

**Fig. 4 |** Dislocation motion via nanoscale segment detrapping mechanism. **a**, Correlated nanoscale processes for a leading partial dislocation (white color) in a sample with $T_a$ = 950 K. The applied temperature and shear stress are 300 K and 300 MPa, respectively. The swept areas between two neighboring snapshots/instants when the dislocation settles down briefly without motion (e.g., 20 ps and 60 ps) are highlighted in red. **b-e**, The calculated minimum energy path of a nanoscale segment detrapping process, for a sample with $T_a$ = 950 K and subject to a local shear stress of 400 MPa. See Methods for calculation details. Reaction coordinates in **b** is the scaled hyperspace arc length. **c** shows the initial configuration with a curved leading partial dislocation (the blue tube). **d** shows the configuration corresponding to the highest saddle point in **b**, and **e** shows the final configuration. For **c**-**e**, surface atoms are colored according to their positions along the axial direction and white atoms represent the complex stacking fault. The nanoscale segment motion is highlighted by the swept area as colored in red, based on the magnitude of atomic displacements along the axial direction. Numbers on the colorbar are in unit of Å.

In order to evaluate the LCO effects on the barriers associated with a typical nanoscale segmented slip process, new samples with smaller dimensions (see Fig. S14 of Supporting Information for size effects on Peierls barrier calculation) are used to calculate the minimum energy path (MEP). Fig. 4b-e shows such an example for a sample with $T_a$ = 950 K (see Methods). As seen in Fig. 4b, a typical nanoscale segment movement traverses quite rugged MEP consisting of multiple finer events with variable barriers, reflecting the complex nature of the underlying energy landscape in concentrated alloys. To complete the entire process (e.g., from Fig. 4c to Fig. 4e), these finer events should be activated in a strongly correlated fashion. Here, to estimate the *effective* barrier associated with the *entire* process, we ignore those finer events whose backward barrier is smaller than the forward barrier. Then the effective barrier is taken as the largest barrier along this modified MEP. For example, the effective barrier (marked with double ended arrow in Fig. 4b) can be taken as the energy difference between the initial configuration (Fig. 4c) and that at the highest saddle point (Fig. 4d). Note that in elemental FCC metals, the Peierls barriers generally vanish when the applied stress is on the order of $10^1$ MPa; however, in the example shown above, the effective barrier is still ~0.15 eV even under a local shear stress of 400 MPa, indicating LCO-induced strengthening.

Next, we systematically compare the effective barriers and the resulting material strengths for samples with various degrees of LCOs. The average values (and standard deviations) of the effective barriers are shown in Fig. 5a, for samples with different LCOs and subject to different stress levels (Methods). See Fig. S15 for a plot encompassing all the effective activation barriers. As seen, a noteworthy trend is that a sample with stronger LCO (or processed at lower $T_a$) tends to impose larger barriers to the nanoscale segment slip process. If fed into the rate equation, these increasingly larger barriers would trap the local dislocation segment for an exponentially increasing time, thus reducing the dislocation mobility. Meanwhile, with increasing shear stresses, the average barriers for all types of samples decrease and eventually vanish at the (average) athermal stress limit, as shown by the fitted curves. Again, samples with stronger LCO show higher (average) athermal stress limit. The complex nature of the underlying energy landscape and of the nanoscale heterogeneities are also directly reflected by the significant standard deviations for each average value. The 300 K average activation volume is shown in Fig. 5b (see Supporting Information for the evaluation of activation volume). Overall, the activation volume associated with the NSD process at normal shear stress levels is in the range of $10^1b^3 – 10^2b^3$, consistent with some recent experimental measurement[14,16] and the direct MD observations shown in Fig. 4a. These relatively small activation volumes suggest a thermally activated process much more sensitive to temperature and strain rate than conventional FCC metals where the activation volumes are much larger.

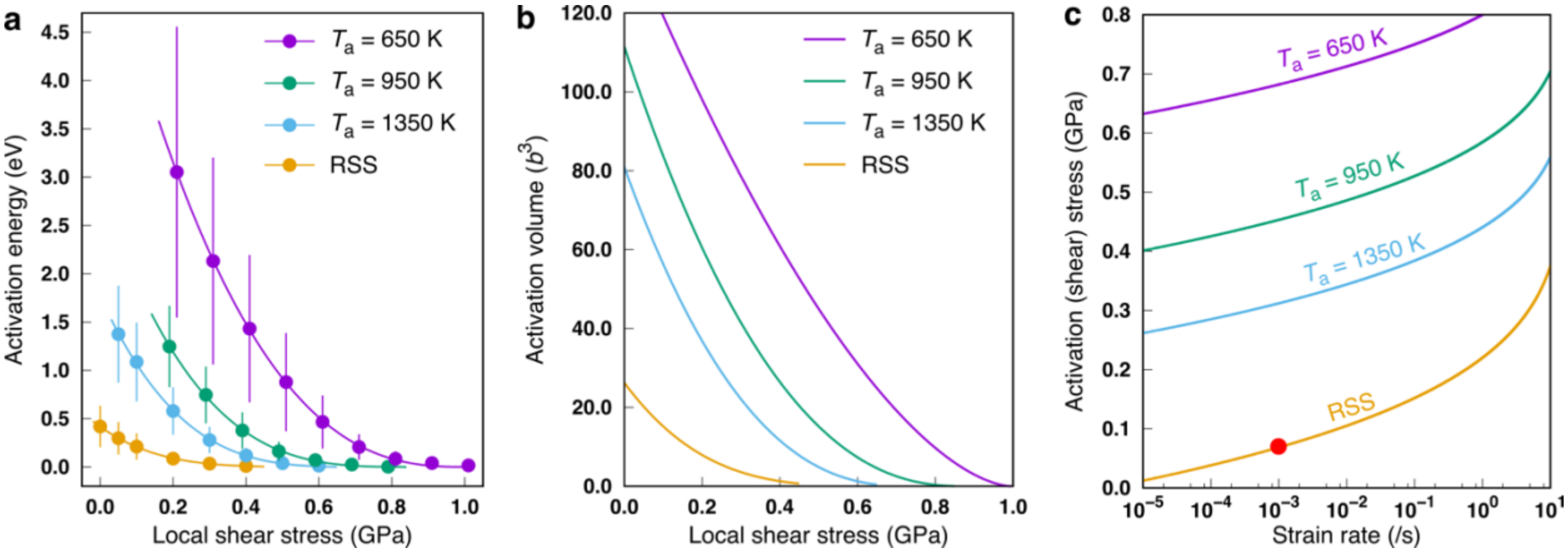

**Fig. 5** | LCO-induced strengthening. **a**, The average activation barriers associated with the nanoscale segment detrapping process. For visualization, data for $T_a$ = 950 K and $T_a$ = 650 K are horizontally shifted by -0.01 GPa and 0.01 GPa, respectively. The calculated data points are fitted using $Q(\tau) = Q_0\left(1 - \frac{\tau}{\tau_{ath}}\right)^{\alpha}$, where $Q_0$, $\tau_{ath}$ and $\alpha$ are fitting parameters. **b**, The average activation volume at 300 K (see Supporting Information for calculation details). **c**, The average activation (shear) stress at 300 K as a function of strain rate. These average activation (shear) stress can be considered as the intrinsic shear strength, which for RSS is ~70 MPa (red point) at $10^{-3}$ $s^{-1}$ (comparable with experiments). With increasing LCO (or decreasing $T_a$), the intrinsic shear strength is predicted to increase, demonstrating LCO-induced strengthening.

The intrinsic shear strength can then be evaluated by solving Orowan's equation (see Supporting Information), and the results are shown in Fig. 5c for different samples at 300 K. As seen, at a given strain rate, the shear stress required to activate NSD increases considerably with increasing LCO (or decreasing $T_a$), clearly demonstrating the pronounced strengthening due to LCO. Such LCO-induced strengthening was indeed observed in experiments on an BCC HEA where the yield strength was increased by 76% after one-day annealing of the as-cast state, which was attributed to the development of short-range-clustering[29]. Here the activation shear stress is taken as the intrinsic shear strength of the MEA lattice, because the mobile dislocation density was assumed to remain on the level for well-annealed samples ($10^8$ /$m^2$, see Section 9 in SI) and extrinsic (dislocation) interactions have not yet kicked in. Since HEAs/MEAs have been assumed to be RSS in many experimental and computational works, here we estimate the critical resolved shear strength (CRSS) of NiCoCr RSS under experimentally relevant conditions. As shown in Fig. 5c, at 300 K and $10^{-3}$ $s^{-1}$ (as marked by the red point), the CRSS is ~70 MPa. Experimentally measured CRSS is ~55 MPa for single-crystal NiCoCr MEA[56] (*axial* yield strength of polycrystals can be above 270 MPa,[16,28,30,45] but there some grain size strengthening is likely involved).

We also observed similar LCO-induced strengthening effects for extended edge dislocations, using MD simulations at 300 K under a constant strain rate, see Fig. S16 - S18 of Supporting Information. The shear stress required to drive forward a leading partial dislocation needs to increase by as much as a factor of three, when an RSS has been annealed at 650 K to increase LCO. Also see Movies S1-S2 for the dynamic motion processes with dislocations evolving into rugged morphologies, detrapping from the nanoscale LCO heterogeneities.

Note that the observed strengthening effect is not due to the lattice distortions caused by atomic size mismatch, as our MD/MC aging at $T_a$ alleviates the effects of atomic size mismatch through rearranging unfavorable local atomic environments. Two mechanisms are believed to be responsible for the additional strengthening due to LCO. First, the spatially heterogeneous complex stacking fault energy and antiphase boundary energy results in extra restoring forces on a moving dislocation that breaks LCOs. The stronger the LCO, the larger restoring force a dislocation feels on average. Second, the spatially varying LCO presents nanoscale heterogeneities acting as trapping roadblocks to moving dislocations, in a way similar to G-P zones. In comparison, in RSS without a dominant LCO, the dislocation still shows a wavy line and its forward motion would still undergo NSD (see Fig. 6), such that the RSS is already strengthened compared to elemental FCC metals. However, in the random solution the NSD only need to detrap from local favorable environment formed due to statistical fluctuation, such as spatial deviations from chemical disorder (see Fig. 6). Such strengthening behavior is consistent with the Labusch theory[57] where strengthening is attributed to the collective interactions between many solute atoms and a dislocation, i.e., it is the favored statistical fluctuations in solute configuration that pin dislocations, rather than individual solute atoms. Therefore, when a dominant LCO emerges in a sample, the activation barriers for local dislocation segments would increase as shown in Fig. 5, the dislocation lines become increasingly wavy due to partially ordered spatial heterogeneities, and the critical resolved shear stress rises further. As a practical approach to take advantage of this LCO strengthening, one can rapidly cool an HEA to obtain an RSS to make use of its room-temperature ductility for shaping, and afterwards age it at an elevated temperature to acquire adequate LCOs to raise its strength for service (e.g., when high strength is needed for use at elevated temperatures).

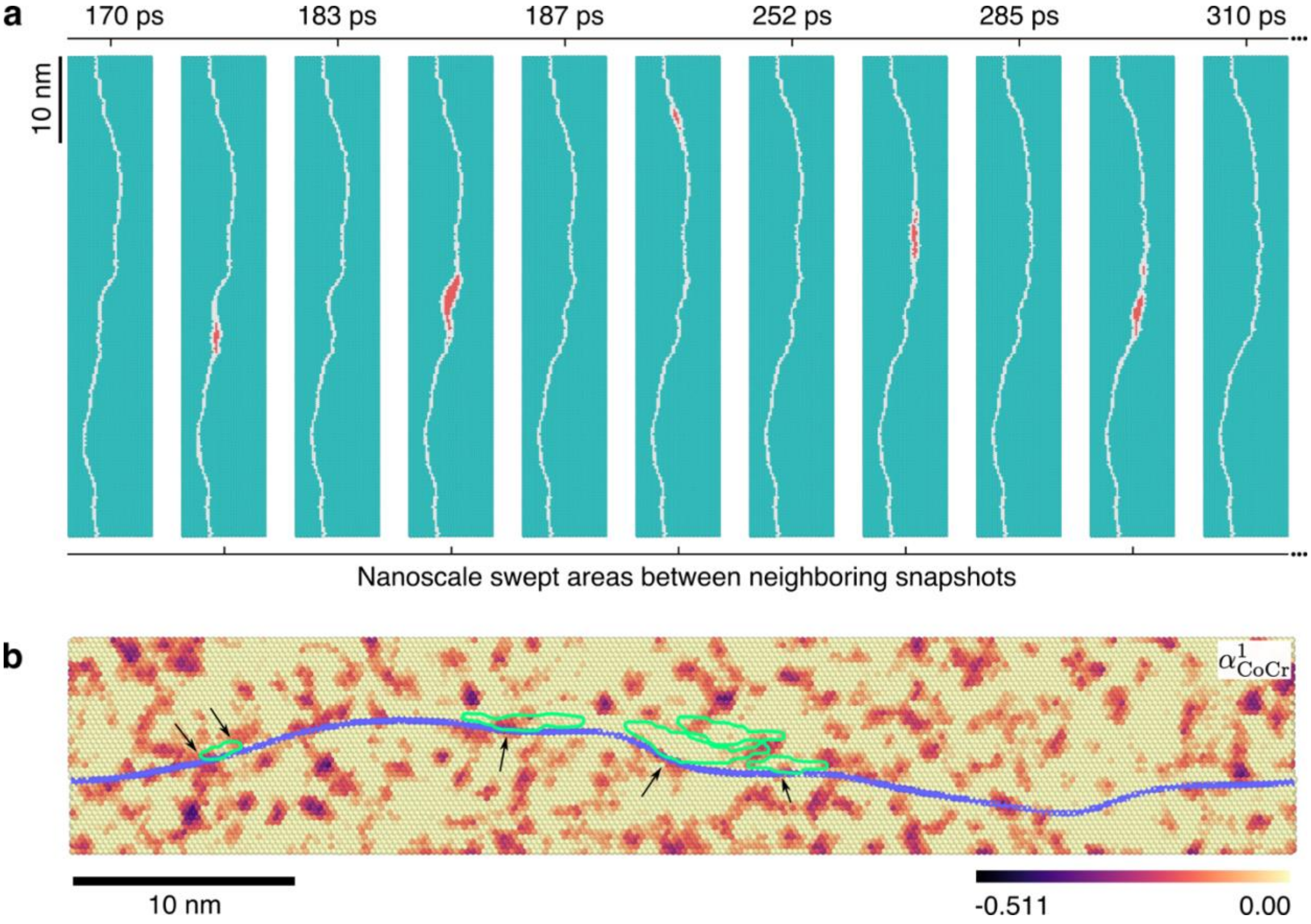

**Fig. 6** | Nanoscale segment detrapping in random solid solution of NiCoCr. **a**, Dislocation line morphology evolution under a shear stress of 50 MPa at 300 K. The several swept areas are highlighted in red. **b**, Correlation between nanoscale segment detrapping events and spatial deviations from chemical disorder. Atoms are colored according to the coarse-grained $\alpha^1_{\mathrm{CoCr}}$ ; darker color means stronger Co-Cr short-range-order (more negative $\alpha^1_{\mathrm{CoCr}}$). The swept areas in **a** are now outlined in green. These activated nanoscale segments are seen to detrap from some of the local hard regions that have stronger Co-Cr LCO (as denoted by the arrows) and then propagate into local regions without such chemical order. It is not expected that our specific loading conditions within the short simulation period would be able to activate all the segments at LCO trapping sites. For some of the regions with weak Co-Cr LCO along the dislocation line, thermal activation is expected to have already moved those segments, producing a wavy dislocation line before loading, when the sample was first relaxed at 300 K for 100 ps. Continued bow-out increasing curvatures is limited by the line tension.

**Summary**

In summary, generally the HEAs are not ideal or regular SSs, and do not have a high configurational entropy close to random mixing as depicted in the literature. The most probable structural state contains partial chemical order. The configurational entropy is reduced from ideal solutions, even for HEAs prepared at rather high temperatures. They are nevertheless still of "high entropy" compared to the ground state traditional SSs and intermetallic compounds, arising from the variability and vastness in the internal LCO configurations when multiple chemical species are alloyed in concentrated proportions.

The significance lies in the realization that different from traditional SSs where the (dilute) solutes are approximately random, delivering almost invariable properties independent of processing, here the variable chemical order can be selected through judicious alloy processing, opening up rich possibilities, including ordering vs segregation, degree of LCO, species involved in ordering, length scale reached through the first, second, and third nearest-neighbor layers to even nanometer domains, and spatial variation across the sample, all in a single-phase SS, even when the LCOs are difficult to resolve and quantify in diffraction experiments.

A specific example we have focused on is that the concentrated SS with LCOs opens a new playground for dislocations, including important benefits. First, the variable and partial chemical order explain on atomic scale the multitude of possible dislocation mechanisms: we have quantitatively mapped out the complex energy landscape, as well as the activation barriers on the rugged kinetic pathways, from the standpoint of the moving dislocation. A consequence is that, depending on how the sample is prepared, the preferred deformation option can be one, or a subset, on the selection menu, including extended dislocations, stacking faults, deformation twining, or martensitic transformation. Second, the variability above explains why macroscopically measured mechanical behavior such as strength, strain hardening and ductility can seem inconsistent at a given overall composition, when the same HEA is processed differently to reach different levels of sample-average LCOs. Third, the LCO variety can explain the observation of spatial property variations from one local region to another, in a given sample. An example is the dislocation splitting width and local SFE reported in experiments[54]. Fourth, the added dimension of LCO control offers a new knob to turn, to enable unprecedented property tuning in a single-phase solution alloy, such as that demonstrated in Ref. [29].

We emphasize that this flexibility afforded by HEA is inaccessible by its counterpart extremes: on the one end is the singular chemically disordered RSS, and the other is the ground-state phases (terminal SSs and intermetallics that are already commonly known), both offering limited property choices. In general, existing engineering alloys typically contain some ordered phases; now in concentrated single-phase HEAs their equivalent is the LCOs. This hinges on, of course, that LCOs can bring drastic changes to dislocation properties, which has indeed been demonstrated explicitly and quantitatively in our discussions above in connection with related experimental observations. Specifically, HEAs show spatially heterogeneous complex stacking faults that are neither the simple intrinsic stacking fault associated with RSS (either concentrated or dilute), nor the long-range ordered complex stacking faults in intermetallics (superlattices). HEAs demonstrate spatially varying APBE, unlike RSS (no APBE) or intermetallics (spatially uniform APBE). Furthermore, for a specific material property while an RSS (or intermetallic) always gives a unique value, an HEA can be made to access different values, depending on how the sample is processed. In addition, DT and MT in HEAs require significantly higher energy input than in random SS, as a result of breaking LCOs layer by layer. Finally, HEAs with increasing LCOs are expected to exhibit strengthening, because the moving dislocations now have to undergo nanoscale segment detrapping to traverse local heterogeneities that impose escalating activation barriers with increasing LCO. These rich structural features and dislocation energetics/dynamics answer the fundamental materials science question as to what is new to the HEAs to distinguish them from the terminal SS and intermetallic compounds. In sum, the emerging HEAs open opportunities for structure-property design through processing to tune LCOs: "there is plenty of room at the bottom" above and beyond what is available with traditional solid solution alloys.

**Acknowledgement**

This work was supported at JHU by NSF-DMR-1804320. Q.J.L. and E.M. acknowledge the computational resources at Maryland Advanced Research Computing Center (MARCC). Part of the computations were also supported by the Texas Advanced Computing Center (TACC) at The University of Texas at Austin. The work at GMU was supported by the NSF under Grant No. DMR-1611064. The authors also thank W. Curtin and W.G. Nöhring for helpful feedback during the EAM potential development.

## Methods

**Empirical interatomic potential development.** An empirical potential for nonmagnetic NiCoCr has been developed in the formalism of embedded-atom-method (EAM)[58–60], by matching a large ab initio database established for the ternary system. The ab initio database includes a large set of atomic configurations with corresponding cohesive energies, atomic forces, and stress tensors. A similar force-matching method has previously been employed to develop a highly optimized potential for the Zr-Cu-Al system[60]. In this work, our focus is on the properties of Ni-Co-Cr entropy alloys, and to that end, special attention has been paid to the energetics of the stacking faults and chemical ordering of Ni-Co-Cr solid solutions.

In order to give an accurate account of the Ni-Co-Cr system in the full compositional range, more than 3,000 atomic configurations were selected to build a comprehensive ab initio database. The atomic configurations not only encompass all the intermetallic compounds reported in the Ni-Co-Cr system, but also include liquid/glass structures, various types of defects, transition pathways, etc. covered in a large pressure-temperature phase space of Ni-Co-Cr. The revised Potfit code[59] was used for potential fitting. Lastly, the potential was improved through an iterative process and was further refined to match experimental data including cohesive energies, lattice parameters, elastic constants and phonon frequencies of the constituent elements. For more details of constructing the database and potential fitting methodology, the readers may refer to refs[60,61].

All ab initio calculations were performed with the density-functional based Vienna Ab-initio Simulation Package (VASP)[62]. We used the projector augmented-wave (PAW) method[63,64] to describe the electron-ion interactions and the generalized gradient approximation (GGA) for exchange-correlation functionals. The valance electrons of Ni, Co, Cr were specified as $3d^8 4s^2$, $3d^7 4s^2$, and $3d^5 4s^2$, respectively. The spin-polarization effect was not considered in the potential fitting (see Supplementary Information for a discussion on magnetic effects). For high-precision ab initio total-energy calculations, we typically used 3x3x3 Monkhorst-Pack[65] k-point grids with each atomic configuration containing 100-200 atoms.

**Molecular dynamics (MD) and Monte Carlo (MC) simulation.** The chemical potential differences with regard to Ni were determined by hybrid MD and MC simulations under the semi-grand canonical ensemble at 1500 K. The set of parameters that minimizes the composition errors with respect to the equiatomic concentration are: $\Delta\mu_{\mathrm{Ni-Co}} = 0.021$ eV and $\Delta\mu_{\mathrm{Ni-Cr}} = -0.31$ eV. This ensures the miscibility of all elements near the equiatomic compositions. Then hybrid MD and MC simulations, under the variance-constrained semi-grand-canonical (VC-SGC) ensemble[66], were carried out to obtain the equilibrium configurations at different annealing temperatures $T_a$. The variance parameter $\kappa$ used in our simulations is $10^3$. Our samples consist of $N$ = 1584000 atomic sites with $x$, $y$ and $z$ along the $[11\bar{2}]$, $[111]$ and $[1\bar{1}0]$ crystal directions, respectively. The dimension of the simulation box are ~ 52 nm × 6 nm × 52 nm. Periodic boundary conditions were applied in all directions. Every 20 MD steps, there is one MC cycle consists of $N/4$ trial moves. The MD timestep was set to 2.5 fs. A Nose-Hover thermostat and a Parrinello-Rahmann barostat were used to control temperature and pressure, respectively. Sufficient MC cycles (270,000 to 500,000 cycles depending on annealing temperature) were carried out to achieve converged LCO. The converged configurations were then quenched to zero temperature and energy-minimized to eliminate thermal uncertainties for subsequent property calculations (e.g., LCO and various fault energies). All simulations were carried out using the LAMMPS package[67] and the atomic configurations were visualized with the Ovito package[68].

**Chemical short-range-order parameters**. The pairwise multicomponent short-range order parameter is defined as $\alpha_{ij}^{m} = \left(p_{ij}^{m} - C_j\right)/\left(\delta_{ij} - C_j\right)$, where $m$ means the $m^{\text{th}}$ nearest neighbor shell of the central atom $i$, $p_{ij}^{m}$ is the average probability of finding a *j*-type atom around an $i$-type atom in the $m^{\text{th}}$ shell, $C_j$ is the average concentration of *j*-type atom in the system and $\delta_{ij}$ is the Kronecker delta function. For pairs of the same species (i.e., $i$ = *j*), a positive $\alpha_{ij}^{m}$ suggests the tendency of segregation in the $m^{\text{th}}$ shell and a negative $\alpha_{ij}^{m}$ means the opposite. In contrast, for pairs of different elements (i.e., $i \neq j$), a negative $\alpha_{ij}^{m}$ suggests the tendency of *j*-type clustering in the $m^{\text{th}}$ shell of an *i*-type atom while a positive $\alpha_{ij}^{m}$ means the opposite. For a specific $T_a$, $\alpha_{ij}^{m}$ are the average values over a series of converged/equilibrium configurations.

**Energy pathway calculations**. All energy pathways were calculated using the equilibrium configurations obtained from the hybrid MD and MC simulation. The as-prepared samples have 30 (111) layers in the $y$ direction; however, we doubled the $y$ direction box length by replicating the sample along [111] direction. Then two neighboring (111) planes were chosen and centered in the simulation box. The simulation box was then divided into two slabs by a cutting plane dividing the two chosen (111) planes. Afterwards, the boundary conditions along [111] was switched to free surface. Complex stacking faults were created by relatively displacing the two slabs according to the Burgers vector of partial dislocations in all three <112> directions. Local antiphase boundaries can then be created by relatively displacing the two slabs according to the appropriated Burgers vectors of trailing partial dislocations. For deformation twinning and martensitic transformation, samples with a complex stacking fault were used as the starting configurations. Deformation twinning nucleation and twin thickening were realized by repeated partial dislocation slipping on consecutive (111) planes, while martensitic transformation was carried out by repeated partial dislocation slip on every other (111) plane. Energy pathways were obtained by constrained energy minimizations on intermediate configurations (only allowed to relax along the [111] direction) linearly interpolated between the initial and the final configurations (both initial and final configurations were fully relaxed). For each $T_a$, the sample-average values of various fault energies were computed based on 90 different energy pathways (30 layers each with three different Burgers vectors). For calculations on the local complex stacking fault energies and local antiphase boundary energies, we divided the whole sample into columns along the [111] direction and each column shares the same cross-sectional area of 3.2 $nm^2$. Then the local stacking fault energies are calculated by considering the potential energy changes of these columns. This method leads to an average value same as the global value. For better visualization in Fig. 3b and Fig. 3d, linear interpolation is carried based on the coarse-grained value (over the first nearest neighbors) for each local area.

**Activation barrier calculations for nanoscale segment detrapping events.** In order to examine the LCO effects on local dislocation segment activation barriers, hybrid MD and MC simulations were performed on relatively smaller samples to obtain equilibrium LCOs at $T_a$ = 650 K, 950 K, and 1350 K, respectively. For meaningful statistics, 30 different samples were obtained for each processing condition. The dimensions for these samples are 20 nm × 20 nm × 10 nm, along $x[1\bar{1}0]$, $y[\bar{1}\bar{1}\bar{1}]$ and $z[11\bar{2}]$, respectively. Similarly, after the MD/MC simulations, samples were quenched to 0 K and energy minimized. Then cylinder configurations with axial direction along $[11\bar{2}]$ and a diameter of 20 nm were cut from the fully relax bulk samples. For each cylinder sample, a partial dislocation with screw character (Burgers vector is along $[11\bar{2}]$) was introduced at the center, based on the anisotropic elasticity theory. The dislocation line length (i.e., the axial length of the cylinder sample) is 10 nm, as the observed detrapping processes are on nanoscale. The outer layers of atoms with a thickness larger than two times the potential cutoff radius were fixed during the subsequent energy minimization. An inner cylindrical region with a diameter of 10 nm was used to calculate the average local shear stress. To create a final state for the minimum energy path calculations, quasi-static athermal shearing (i.e., a small uniform shear strain + energy minimization at each step) was employed to trigger events at the athermal stress limit. Then the configuration with new events was unloaded to desired strain levels, serving as the final states. Minimum energy paths were then calculated using the simplified and improved string method[69]. FIRE algorithm[70] was used to update each image during the evolution step and linear interpolation was used for the parameterization. Reparameterization was carried out every 10 increments of the minimizer. We stop the iteration once the displacement of every image between two consecutive iterations is less than $10^{-3}$ Å or after a total number of 3,000 iterations. We have double checked that the calculated minimum energy paths are comparable to well-converged nudged elastic band calculations (see Section S11 in Supporting Information for verification).

Supporting Information

# Strengthening in multi-principal element alloys with local-chemical-order roughened dislocation pathways

Qing-Jie Li[1], Howard Sheng[2,3*], Evan Ma[1*]

[1]Department of Materials Science and Engineering, Johns Hopkins University, Baltimore, MD 21218, USA;

[2]Department of Physics and Astronomy, George Mason University, Fairfax, Virginia 22030, USA.

[3]Center for High Pressure Science and Technology Advanced Research, Shanghai 201203, China

[*]To whom correspondence may be addressed. Email: hsheng@gmu.edu; ema@jhu.edu

This file contains the following sections:

Movie S1: Dynamic motion of an extended edge dislocation in a sample processed at $T_a$ = 1350 K.

Movie S2: Dynamic motion of an extended edge dislocation in a sample prepared at $T_a$ = 650 K.

## S1. Validation of the new empirical interatomic potential for NiCoCr

The predicted physical properties from our newly developed NiCrCo EAM (embedded atom method) potential are compared with the experimental/DFT values in Table S1-S2 and Fig. S1-S4. Very good performance has been achieved for pure elements, intermetallic compounds and complex solid solutions. More discussions on magnetic effects will be presented in Section S12.

**Table S1 |** EAM potential predicted properties for Ni, Co, Cr, in comparison with experiments (or ab initio calculations), where $E_c$ is the cohesive energy, $C_{11}$, $C_{12}$ and $C_{44}$ elastic constants, $\nu$ phonon frequencies.

| **Ni** | **EAM** | **Experiment / Theory** |
|---|---|---|
| $E_c$ (eV/atom): fcc, 300K, $a$ = 3.52 Å | -4.45 | -4.45[a] |
| $C_{11}$ (GPa) | 262 | 262[b] |
| $C_{12}$ (GPa) | 155 | 150[b] |
| $C_{44}$ (GPa) | 122 | 131[b] |
| $\nu_L(X)$ (THz) (fcc $a$ = 3.52 Å) | 8.57 | 8.55[c] |
| $\nu_T(X)$ (THz) (fcc $a$ = 3.52 Å) | 6.18 | 6.27[c] |
| $\Delta E_{\mathrm{fcc-hcp}}$ (eV/atom) | 0.03 | 0.03[d] |
| $\Delta E_{\mathrm{fcc-bcc}}$ (eV/atom) | 0.10 | 0.09[d] |
| **Co** | **EAM** | **Experiment / Theory** |
| $E_c$ (eV/atom): hcp, 300K, $a$ = 2.507 Å, $c$ = 4.069 Å | -4.39 | -4.39[a] |
| $C_{11}$ (GPa) (fcc $a$ = 3.53 Å) | 260 | 259[b] |
| $C_{12}$ (GPa) (fcc $a$ = 3.53 Å) | 165 | 159[b] |
| $C_{44}$ (GPa) (fcc $a$ = 3.53 Å) | 102 | 109[b] |
| $\nu_L(X)$ (THz) (fcc $a$ = 3.53 Å) | 8.06 | 8.1[c] |
| $\nu_T(X)$ (THz) (fcc $a$ = 3.53 Å) | 5.75 | 5.8[c] |
| $\Delta E_{\mathrm{fcc-hcp}}$ (eV/atom) | 0.014 | 0.016[d] |
| $\Delta E_{\mathrm{fcc-bcc}}$ (eV/atom) | 0.12 | 0.13[d] |
| **Cr** | **EAM** | **Experiment / Theory** |
| $E_c$ (eV/atom): bcc, 300K, $a$ = 2.91 Å | -4.10 | -4.10[a] |
| $C_{11}$ (GPa) (bcc $a$ = 2.91 Å) | 380 | 391[b] |
| $C_{12}$ (GPa) (bcc $a$ = 2.91 Å) | 161 | 89[b] |
| $C_{44}$ (GPa) (bcc $a$ = 2.91 Å) | 77 | 103[b] |
| $\nu$ ($X$) (THz) (bcc $a$ = 2.91 Å) | 6.5 | 7.8[c] |
| $\nu$ ($P$) (THz) (bcc $a$ = 2.91 Å) | 7.94 | 8.2[c] |
| $\Delta E_{\mathrm{bcc-fcc}}$ (eV/atom) | 0.22 | 0.4[d] |
| $\Delta E_{\mathrm{bcc-hcp}}$ (eV/atom) | 0.22 | 0.46[d] |

[a]Ref. 1; [b]Ref. 2; [c]Ref. 3; [d]ab initio calculation in the present work.

Note: the as-developed Ni-Co-Cr EAM potential is available upon request.

Table S1 lists the basic properties of elemental structures calculated from our EAM potential and the corresponding experimental/theoretical values. Fig. S1 shows the equations of state of different allotropes of the elements. The EAM potential can correctly predict the ground states of the elements in accordance with ab initio calculations. For Co, the cohesive energies of face-centered-cubic (FCC) and hexagonal close packed (HCP) structures are very close, with FCC Co being the ground state in the non-spin polarized DFT calculation. Fig. S2 shows the predicted cohesive energies for the three binary systems involved. The energies and lattice constants of the energetically optimized structures are also provided in Table S2. The energy differences between EAM and ab initio calculations (Table S2) are within several tens of meV, which is indicative of a high-quality interatomic potential for metallic alloys.

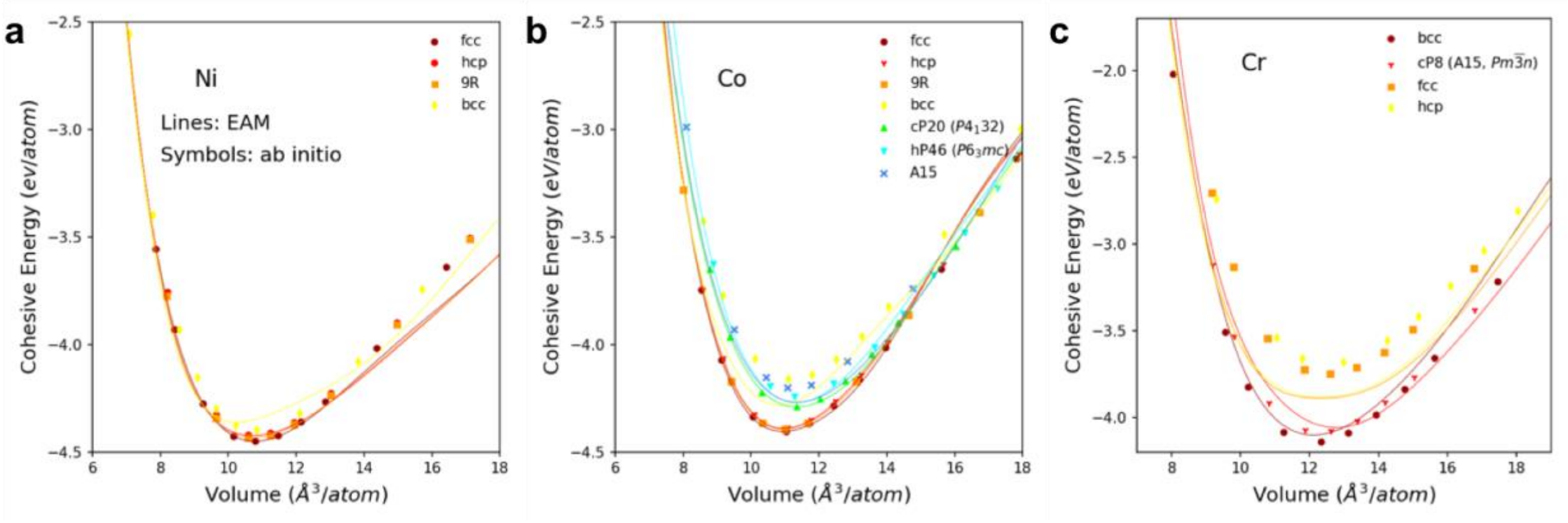

**Fig. S1 |** Equations of state of selected crystal structures. **a**, Equation of state for Ni. **b**, Equation of state for Co. **c**, Equation of state for Cr.

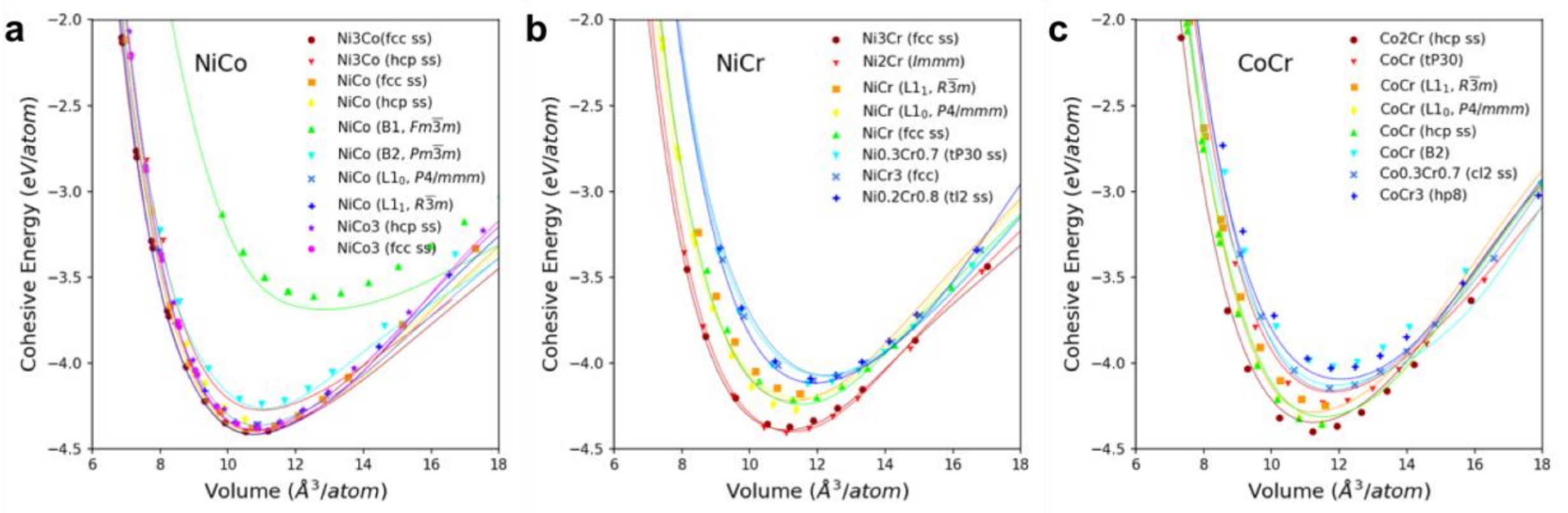

**Fig. S2 |** Comparisons of *ab initio* and EAM calculations of the cohesive energies of different intermetallics. **a**, Ni-Co intermetallics. **b**, Ni-Cr intermetallics. **c**, Co-Cr intermetallics.

**Table S2** | Comparisons between EAM predications and ab initio evaluations of the cohesive energy and lattice parameters of selected intermetallic compounds for the three binary systems. SS: solid solution.

| | | EAM | | Ab initio | |
|---|---|---|---|---|---|
| **Composition** | Structure | Lattice parameter (Å) | $E_c$ (eV/atom) | Lattice parameter (Å) | $E_c$ (eV/atom) |
| **$Ni_3Co$** | FCC SS | 3.506 | -4.419 | 3.510 | -4.407 |
| **NiCo** | FCC SS | 3.514 | -4.402 | 3.516 | -4.385 |
| **NiCo** | B1, $Fm\bar{3}m$ | 3.710 | -3.688 | 3.690 | -3.630 |
| **NiCo** | B2, $Pm\bar{3}m$ | 2.810 | -4.265 | 2.801 | -4.240 |
| **NiCo** | $L1_0$, $P4/mmm$ | $a$ = 3.522<br>$b$ = 3.503 | -4.365 | $a$ = 3.530<br>$b$ = 3.553 | -4.366 |
| **NiCo** | $L1_1$, $R\bar{3}m$ | $v$ = 10.748 Å$^3$ | -4.408 | $v$ = 10.8565 Å$^3$ | -4.374 |
| **$NiCo_3$** | FCC SS | 3.519 | -4.395 | 3.519 | -4.385 |
| **$Ni_2Cr$** | $Immm$ | $a$ = 2.496<br>$b$ = 3.600<br>$c$ = 7.50 | -4.399 | $a$ = 2.484<br>$b$ = 3.583<br>$c$ = 7.463 | -4.404 |
| **$Ni_3Cr$** | FCC SS | 3.546 | -4.390 | 3.553 | -4.383 |
| **NiCr** | $L1_1$, $R\bar{3}m$ | $v$ = 11.292 Å$^3$ | -4.221 | $v$ = 11.520 Å$^3$ | -4.190 |
| **NiCr** | $L1_0$, $P4/mmm$ | $a$ = 3.599<br>$b$ = 3.579 | -4.210 | $a$ = 3.595<br>$b$ = 3.576 | -4.265 |
| **NiCr** | FCC SS | 3.584 | -4.242 | 3.559 | -4.212 |
| **$Ni_{0.3}Cr_{0.7}$** | SS., $P4_2/mnm$ | $a$ = 8.910<br>$b$ = 4.643 | -4.075 | $a$ = 8.682<br>$b$ = 4.530 | -4.118 |
| **$NiCr_3$** | FCC SS | 3.667 | -4.076 | 3.624 | -4.103 |
| **$Ni_{0.2}Cr_{0.8}$** | BCC SS | 2.876 | -4.116 | 2.868 | -4.093 |
| **$Co_2Cr$** | HCP SS | $a$ = 2.525<br>$c$ = 4.065 | -4.347 | $a$ = 2.527<br>$c$ = 4.069 | -4.405 |
| **CoCr** | $L1_1$, $R\bar{3}m$ | $v$ = 11.263 Å$^3$ | -4.290 | $v$ = 11.608 Å$^3$ | -4.261 |
| **CoCr** | SS, tP30, $P4_2/mnm$ | $a$ = 8.785<br>$c$ = 4.553 | -4.172 | $a$ = 8.737<br>$c$ = 4.528 | -4.231 |
| **CoCr** | $L1_0$, $P4/mmm$ | $a$ = 3.587<br>$b$ = 3.568 | -4.316 | $a$ = 3.592<br>$b$ = 3.572 | -4.315 |
| **CoCr** | HCP SS | $a$ = 2.549<br>$c$ = 4.103 | -4.317 | $a$ = 2.543<br>$c$ = 4.094 | -4.362 |
| **CoCr** | B2, $Pm\bar{3}m$ | 2.872 | -4.124 | 2.870 | -4.016 |
| **$Co_{0.3}Cr_{0.7}$** | BCC SS | 2.868 | -4.160 | 2.875 | -4.150 |
| **$CoCr_3$** | hP8, $P6_3/mmc$ | $a$ = 5.183<br>$c$ = 4.158 | -4.094 | $a$ = 5.130<br>$c$ = 4.116 | -4.040 |

Now we consider the performance of our EAM potential on the ternary NiCoCr solid solutions. The as-obtained potential has been tested for a number of physical properties of NiCoCr medium-entropy alloys. We first show the performance of the potential in describing NiCoCr alloys in a large phase space. Fig. S3 shows the comparisons of the energies of NiCoCr solid solutions estimated using EAM and ab initio treatments. As seen, the EAM results are highly consistent with the DFT results, suggesting a high accuracy of our EAM potential.

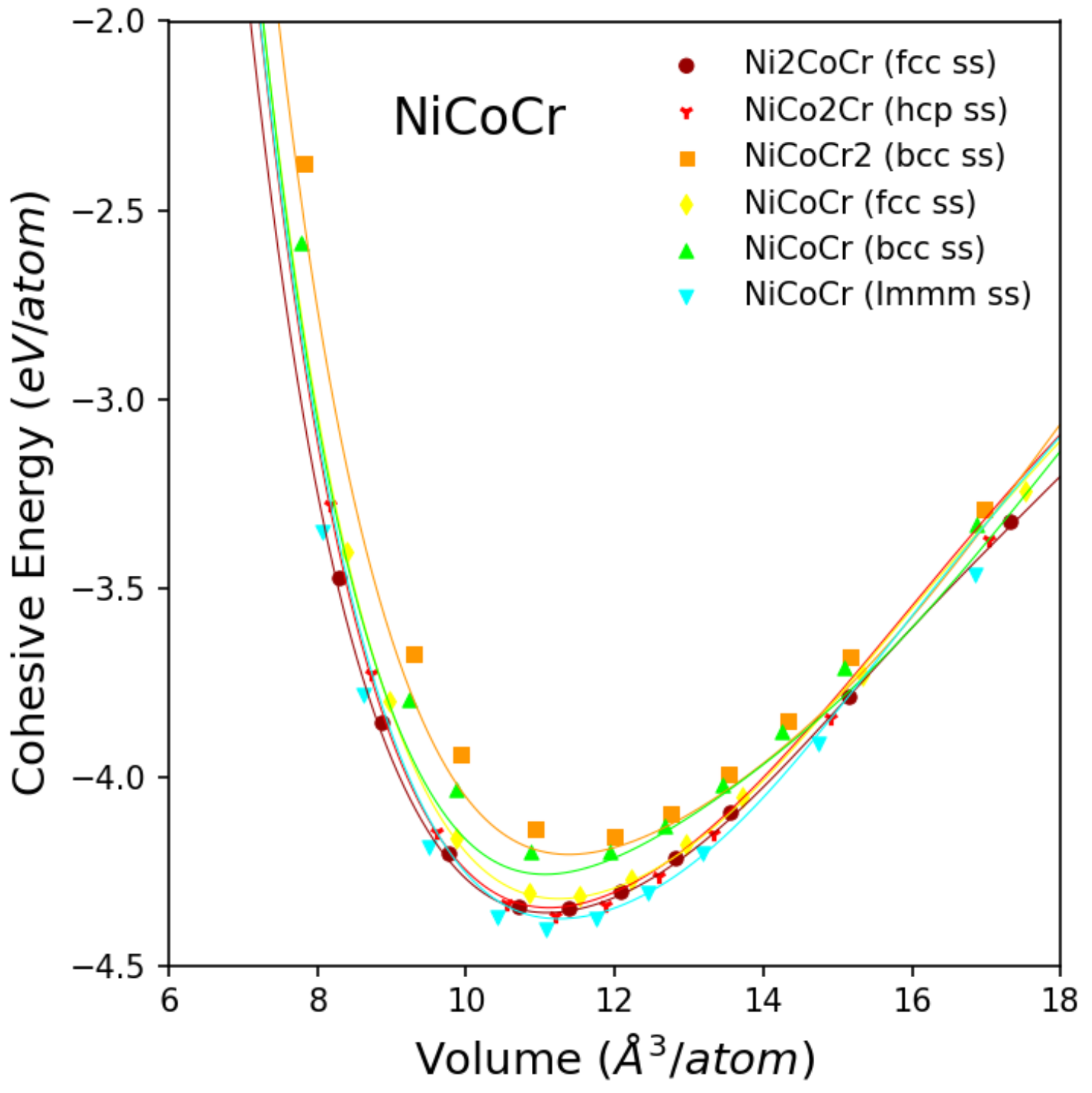

**Fig. S3** | The cohesive energies of selected NiCoCr solid solutions as a function of volume.

The as-developed potential has been utilized to evaluate the physical properties of NiCoCr solid solutions in the entire compositional range, which would otherwise be impossible with the DFT calculations, as shown in Figure S4. During our calculation, a large system containing 72,000 atoms with the desired composition was created with atoms randomly arranged on an FCC lattice, followed by conjugated-gradient energy minimization to find the lowest energy minimum. The elastic constants were analytically calculated based on the second-derivatives of the potential energy. It can be seen that the EAM potential is applicable to study FCC NiCoCr in the entire compositional range, and all the configurations are found to be mechanically stable based on the Born criterion $C_{11} - C_{12} > 0$ and $C_{44} > 0$. In terms of cohesive energy and lattice constant, it is found that both quantities can be described by the rule of mixtures for FCC NiCoCr solid solutions, where the deviations of the cohesive energy and the lattice constant from the rule of mixtures are within 1.2% and 0.3%, respectively.

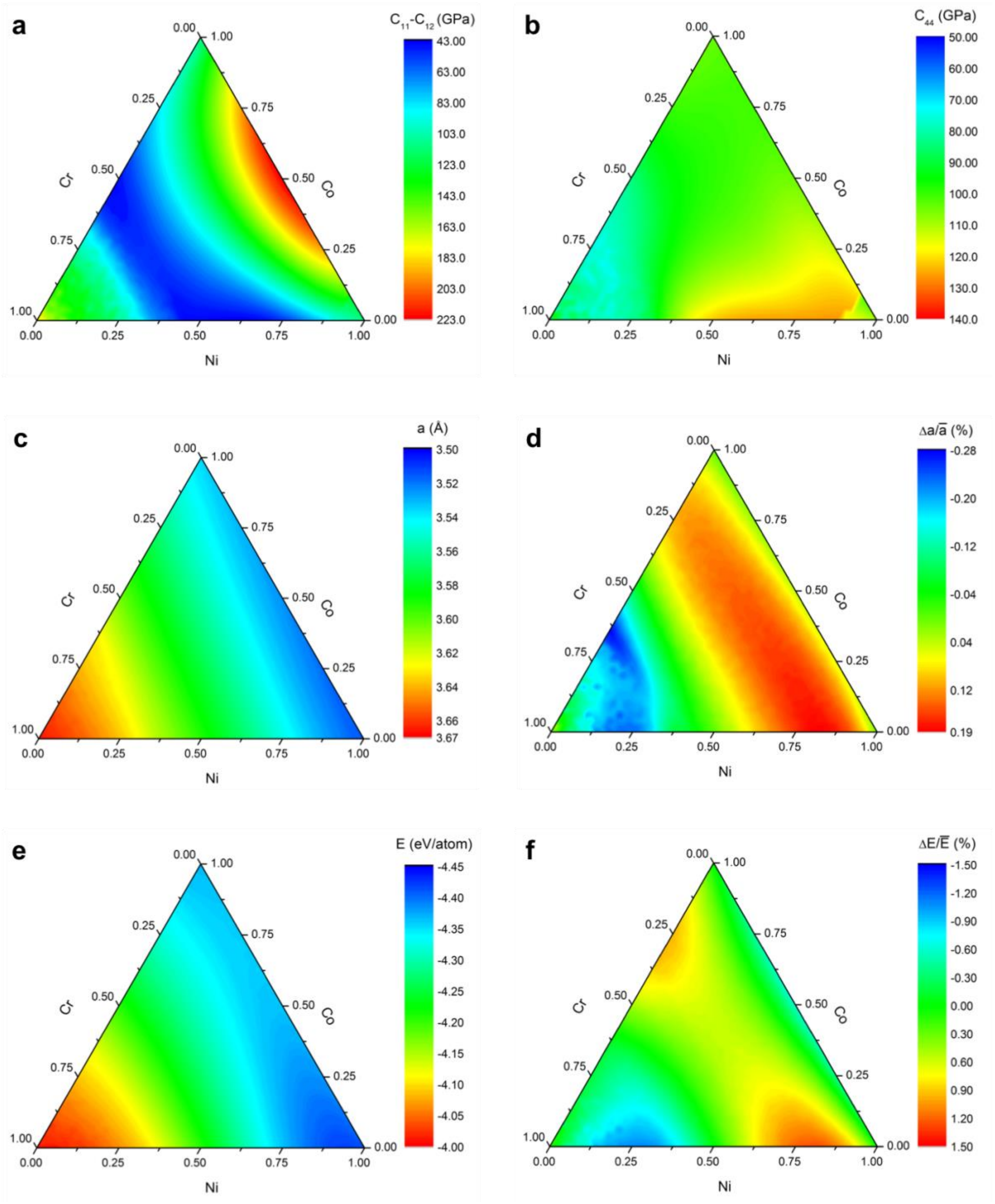

**Fig. S4** | EAM predictions on physical properties of NiCoCr alloys in the entire compositional range. **a-b**, FCC phase of NiCrCo in the entire composition region is mechanically stable, based on the Born criterion $C_{11} - C_{12} > 0$ and $C_{44} > 0$. **c**, lattice constants $a$. **d**, deviations of calculated $a$ from that of rule of mixture. **e**, cohesive energies $E_c$. **f**, deviations of calculated $E_c$ from that of rule of mixture. **c-f** suggest that Vegard's law is obeyed by the Ni-Co-Cr solid solutions, i.e., the $a$ ($E_c$) of an FCC NiCoCr solid solution is approximately equal to the geometric mean of the constituents' a ($E_c$), largely satisfying the rule of mixture.

We also calculated the generalized stacking fault energy (GSFE) of random NiCoCr alloy (Fig. S5), using the newly developed EAM potential. We constructed random solid solutions of equi-atomic NiCoCr alloy, with the [111] direction aligned along the *z*-axis. Periodic boundary conditions were only applied in the *x*, *y* directions (a similar method has been adopted in our previous work[2]). The size of the *x*-*y* plane is set to 12.42 × 12.92 $Å^2$, which is close to the critical size for dislocation nucleation under certain stress and temperature conditions[3]. The upper half of the crystal was displaced along the $[11\bar{2}]$ direction on the (111) slip plane. The GSFE line shape was obtained by conducting statistical GSFE analysis on >10,000 random configurations. The position dependent intensity indicates the probability for a GSFE line to be located. For a specific cross-section along the line, we can also derive the distributions of the fault energies, similar to that shown in Fig. 3 of the main text. For example, the cross-section for intrinsic stacking fault energy suggests a wide distribution of the intrinsic stacking fault energy with a mean negative value, which is consistent with previous work[2,4], further validating the accuracy of our EAM potential.

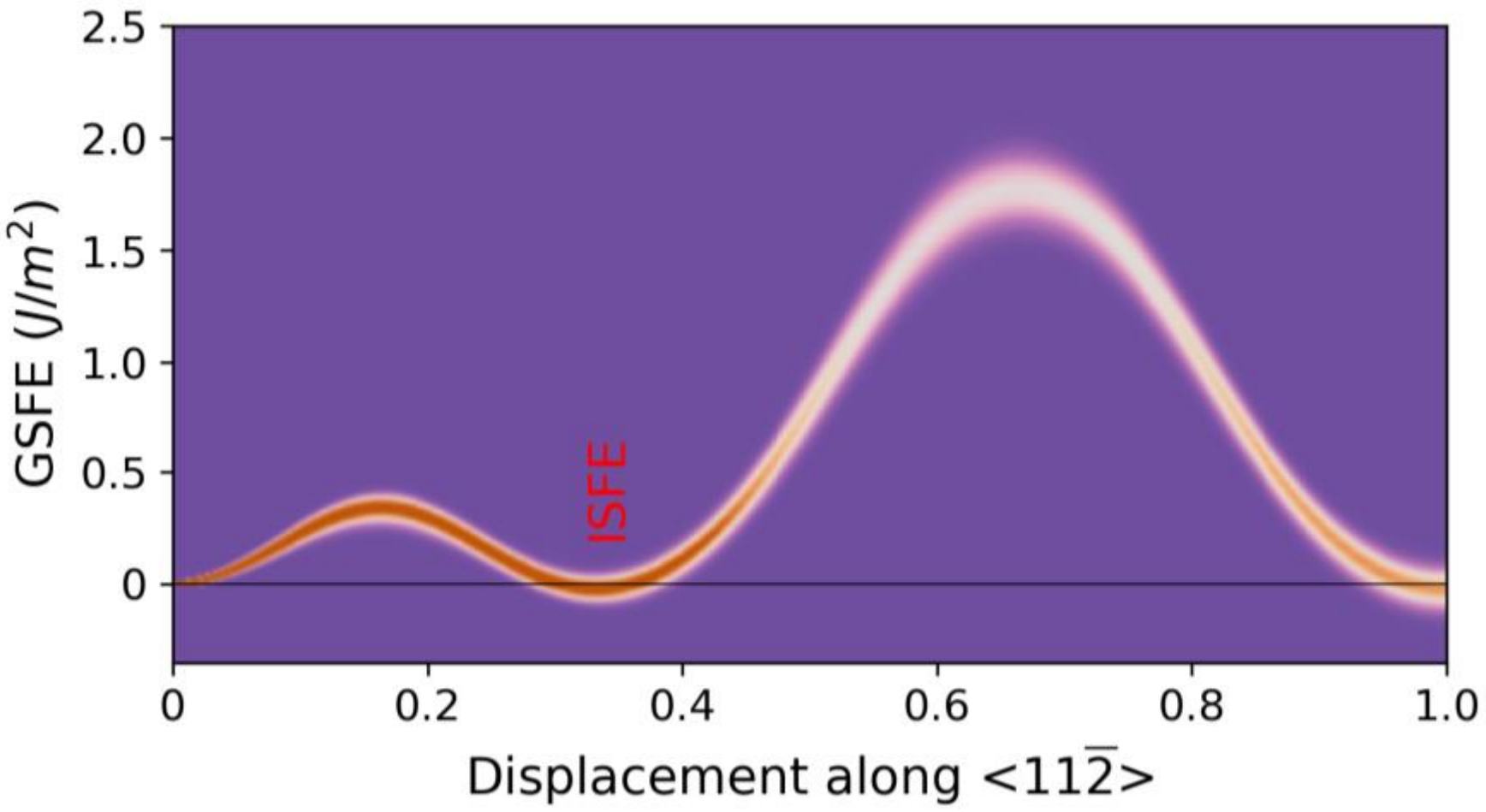

**Fig. S5** | Line broadening of generalized stacking fault energy curve for random solid solutions at 0 K. The stacking fault energy has a finite distribution, which is a hallmark signature of HEAs. The wide distribution of the stacking faults has important implications for the mechanical properties of the chemically disordered crystal. The unit length along the displacement direction is $a\sqrt{6}/2$, where $a$ is the lattice constant.

Within the context of stacking fault energy (SFE), most of the previous studies of HAEs concern the average value of SFE, where the error bars are associated with uncertainties in experimental measurements. Here, by "line broadening", we refer to a wide statistical distribution of the SFE that exhibits a broadened line profile (an intrinsic feature of HAE rather than uncertainties from measurements). It contains more information than an average SFE value plus an error bar. The line profile of the SFE distribution has a significant impact on the deformation behavior of the HAE, as discussed in the main text.

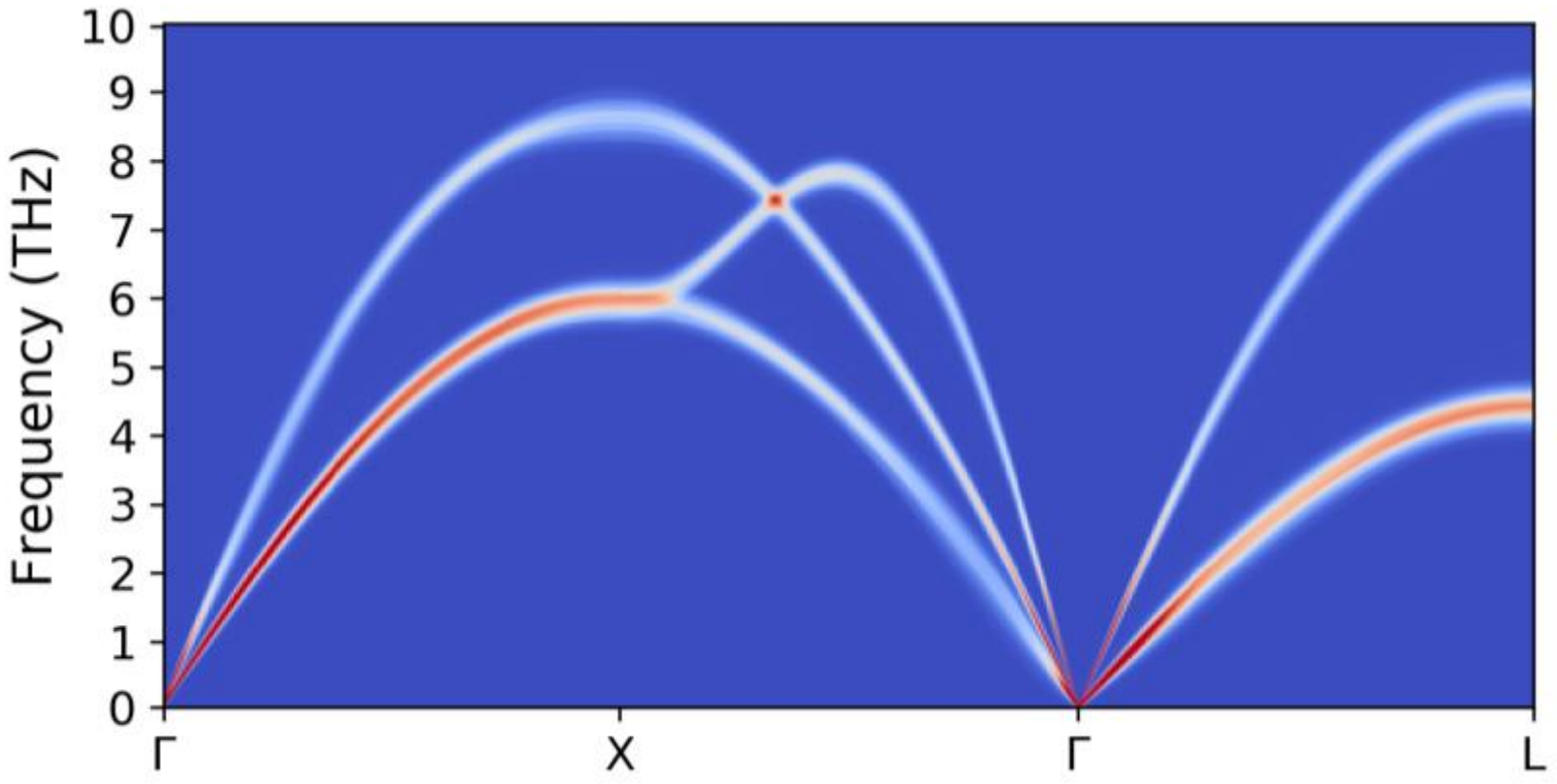

**Fig. S6** | Phonon dispersion line broadening in homogeneously disordered NiCoCr MEA due to mass disorder and force-constant disorder, obtained with the EAM interatomic potential.

Line-broadening of the phonon dispersion curves has been known for homogeneously disordered solid solutions[5,6], such as Fe-Cr[7], due to mass disorder and force-constant disorder. There is also increasing evidence to show that the local chemical order (LCO) has an effect on the lattice dynamics, hence the stability of the crystals[5]. Correct interpretation and prediction of the lattice dynamics of HAE are therefore important. Theoretically, the phonon dispersions of random solid solutions can be treated with different approaches, e.g., the coherent crystal approximation[5] and the special quasi random structures (SQS) method[8]. However, as far as the line shape of the phonon spectra of disordered alloys is concerned, the calculations are much more involved. Here we use a different method to calculate the line broadening of the phonon dispersion curves based on the empirical potential.

For any given lattice parameter and wave vector, the phonons can be computed by diagonalizing the dynamical matrix according to[9]

$$D_{\lambda\mu}(\mathbf{q}) = \frac{1}{\sqrt{m_i m_j}} \sum_{lk} \Phi_{\lambda l \mu k} \ \exp[i\mathbf{q} \cdot (\mathbf{R}_l - \mathbf{R}_k]$$

where $i$ and $j$ are particle indices; $m_i$ and $m_j$ are masses of particle $i$ and $j$; $\alpha$ and $\beta$ are force components ($x$, $y$, or $z$); $\lambda = 3i + \alpha$ and $\mu = 3j + \beta$. The summation over $l, k$ represents the sum over lattice vectors $\mathbf{R_l}, \mathbf{R_k}$ within the cutoff radius. $\Phi_{\lambda l \mu k}$ is the force constant.

In this work, we used the temperature-dependent effective potential method (TDEP)[10] to obtain the effective interatomic force constants (IFCs) of a supercell (108 atoms) of NiCoCr random solid solution. Classical MD simulations were carried out at 300 K to derive atomic displacements and atomic forces. The IFCs were optimized with the *alamode* code[11] based on the force and displacement relationship. Having obtained the IFCs, the dynamical matrix was derived by assuming an averaged atomic mass occupying the FCC unit-cell following the above equation. This method provides a rapid route to map out the phonon dispersion, enabling the estimate of the line shape of the phonon dispersion curves as shown in Fig. S6 (collected over 50,000 different atomic configurations of random NiCoCr). As far as the phonons are concerned, random solid solutions represent a special type of crystal where the lattice periodicity is well defined (this is different from amorphous alloys), as evidenced from the X-ray diffraction patterns of solid solutions. In the reciprocal space, there is a full Brillouin zone associated with the lattice. This is the reason why random solid solutions can be treated with virtual crystal approximation (VCA) in first principles calculations. Experimentally, the phonon dispersion curves can be measured by various techniques such as inelastic x-ray scattering. Phonon line broadening of 12 bcc HAEs was recently investigated with ab initio treatments[6].

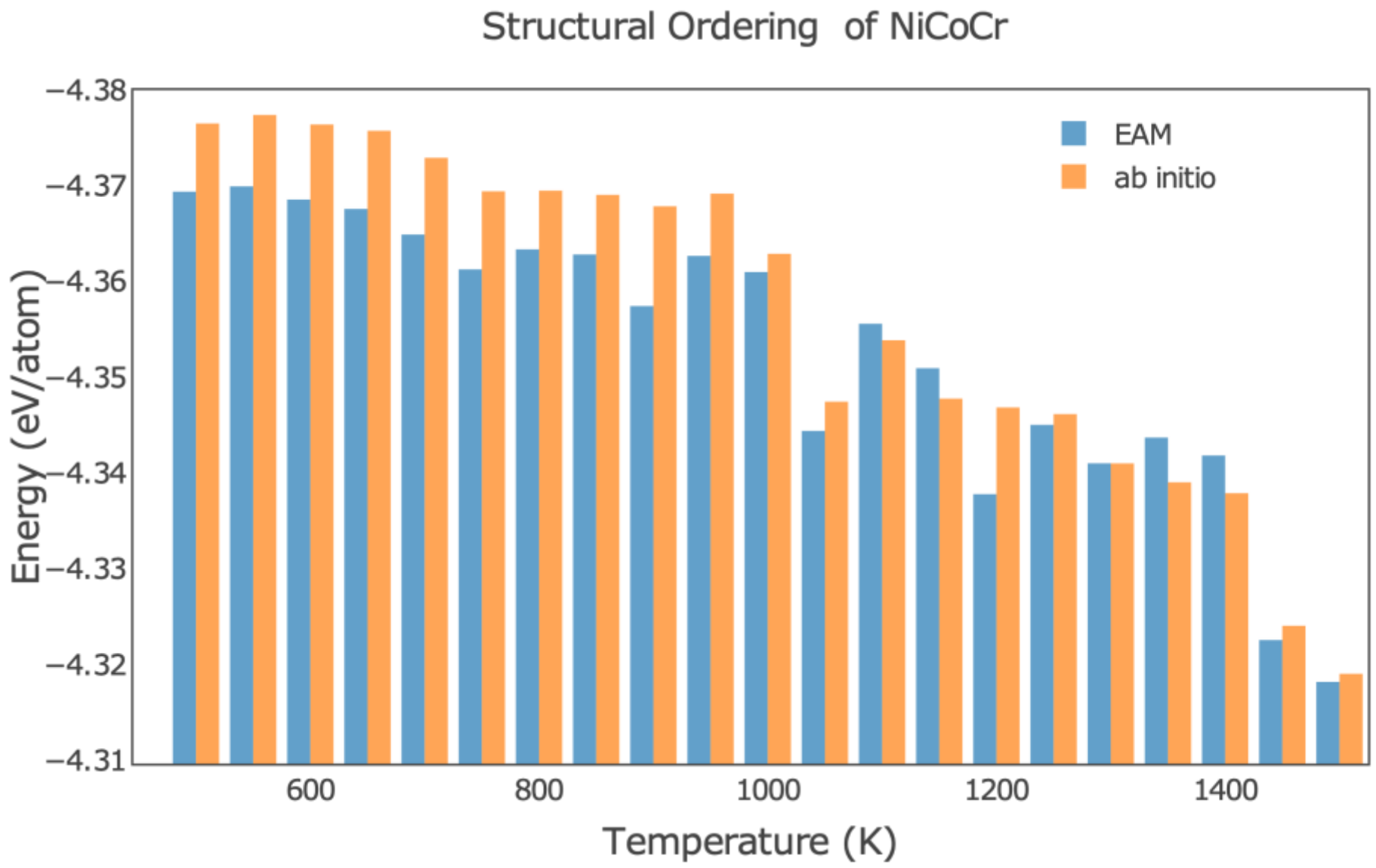

**Fig. S7**. Comparisons of the energies of NiCoCr alloy with different degrees of chemical ordering by EAM and ab initio calculations respectively. The chemical ordering in the alloy was achieved by conducting hybrid MD/MC simulations in a 360-atom ensemble quenched from 1500 K, employing the as-developed empirical EAM potential.

We show in Fig. S7 that the EAM potential yields a cohesive energy trend with increasing chemical ordering, consistent with the DFT calculations. This suggests that our EAM potential is capable of predicting the correct trend of potential energy changes due to chemical ordering during the annealing process.

Chemical ordering is a common phenomenon associated with concentrated solid solutions[12], such as Fe-Cr[13] and Ni-Cr[14]. How to correctly predict and characterize the ordering states in these alloys under different thermodynamic conditions remains a long-standing materials science problem. The chemical ordering of different species is dictated by the thermodynamics of the system, which is a manifestation of interatomic interactions. Hence, the prediction of chemical ordering would require accurate descriptions of the atomic interactions, which, however, is highly challenging for multicomponent systems because many different types of atomic pairs are involved, and the many-body type interactions are complex.  As such, the correct trend of chemical ordering may serve as a reliability test for the validation of the interatomic potentials.

To study the chemical ordering process in the NiCoCr alloy, we are primarily concerned with two questions: (1) whether chemical ordering is going to happen, and (2) what type of chemical ordering is expected if it is kinetically permitted.

Employing the newly developed NiCoCr EAM potential, we carried out hybrid MD/MC simulations to investigate the chemical ordering in equi-atomic NiCoCr alloy cooled down from high temperatures. At each temperature, the atoms in the system were swapped periodically based on the Metropolis algorithm to facilitate atomic diffusion kinetics.  Naturally, the system would evolve toward more thermodynamically stable states with different degrees of chemical order. To better assess the energy states of the atomic configurations, we conducted energy minimization to remove the thermal vibrations. The energies of the as-obtained configurations were subjected to high-precision *ab initio* calculations for cross-checks. The comparisons of the energies of the chemically ordered structures calculated using EAM and ab initio treatments are shown Fig. S7.

It can be seen that the energy trend of "ordered states" predicted by EAM agrees with that from ab initio calculations, that is, the cohesive energy decreases with increasing chemical ordering, suggesting that the as-developed NiCrCo potential is capable of capturing the increasing stabilities with increasing chemical ordering.

## S2. Analysis on the chemical short-range order parameter $\alpha^2$ and $\alpha^3$

The chemical short-range order parameters $\alpha^2$ and $\alpha^3$ for different element-pairs are shown in Fig. S8 for a wide range of annealing temperatures. For the second nearest neighbor shell (SNNS), Ni appears to be segregated at all $T_a$. Co-Co and Cr-Cr chemical ordering experiences an obvious transition around 750 K, i.e., Co-Co pairs and Cr-Cr pairs are not favored in the SNNS above 750 K; but the trend is reversed when $T_a \leq 750$ K, consistent with the formation of more ordered Co-Cr domains shown in Fig. 1 of the main text. The Co-Cr interactions of the SNNS remain similar to those of the first nearest neighbor shell (FNNS), i.e., it tends to form Co-Cr neighbors. Other Ni-related chemical orders remain similar to that for FNNS. For the third nearest neighbor shell (TNNS), the Co-Cr interactions still remain strong, i.e., it tends to form Co-Cr clusters in the TNNS. Co, Cr and Ni all tend to attract the same species in this TNNS.  Ni-Co and Ni-Cr pairs are not favored in the TNNS.

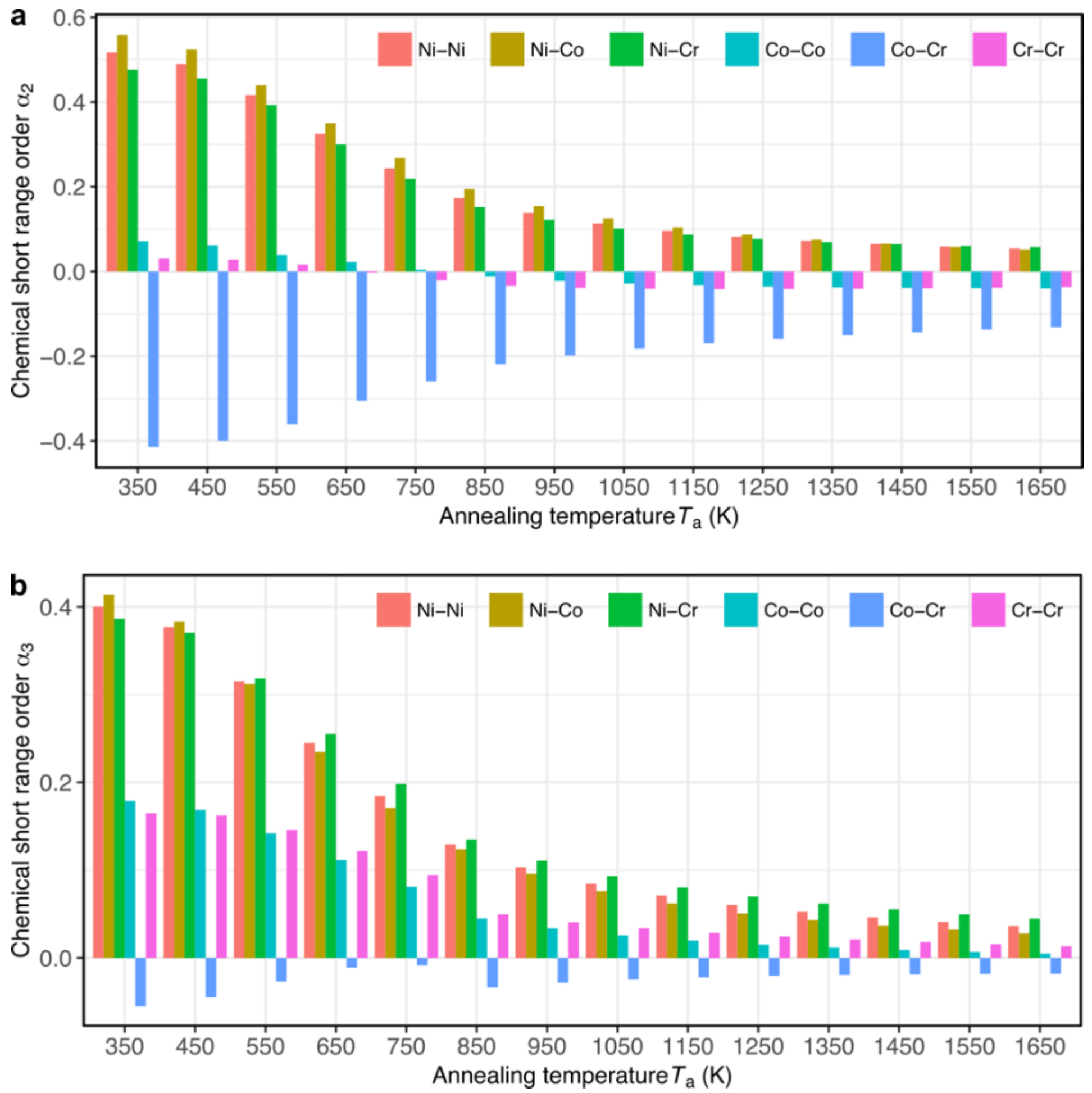

**Fig. S8 |** Chemical short-range order parameter $\alpha^2$ (**a**) and $\alpha^3$ (**b**) at different annealing temperatures.

## S3. Effects of increasing slip amount on complex stacking fault energy, antiphase boundary energy and twin boundary energy

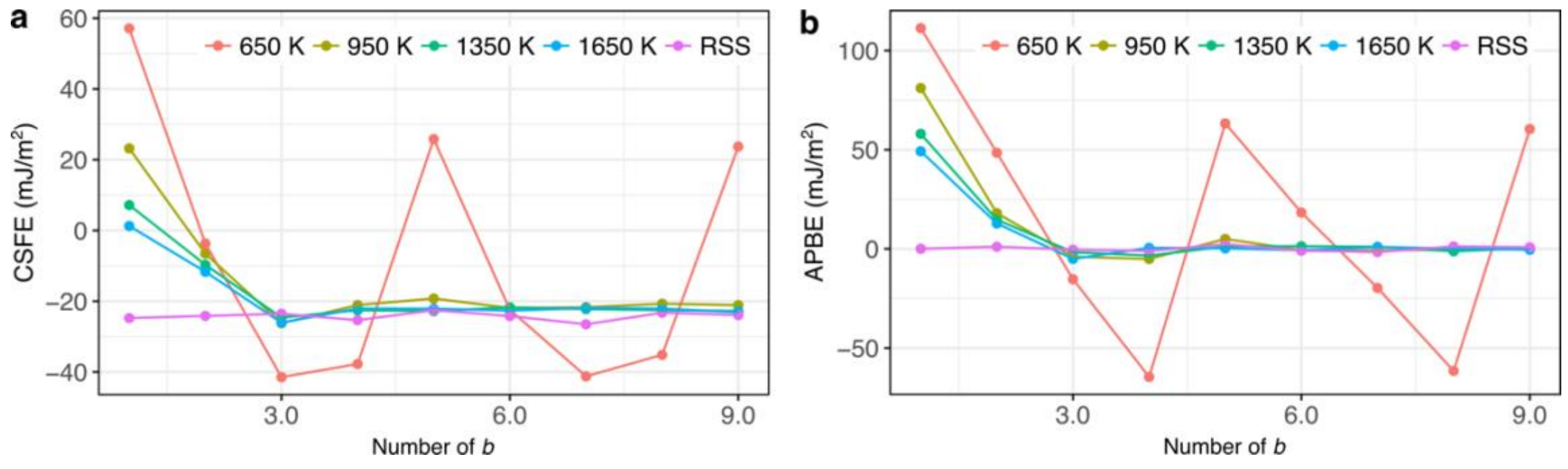

**Fig. S9 |** Effects of increasing slip amount on fault energies. **a**, Variations of complex stacking fault energy with increasing number of Burgers vector *b* on the same slip plane. **b**. Variations of

anti-phase boundary energy with increasing amount of slip. Temperatures in the legends means annealing temperature while all the calculations are carried out at 0 K.

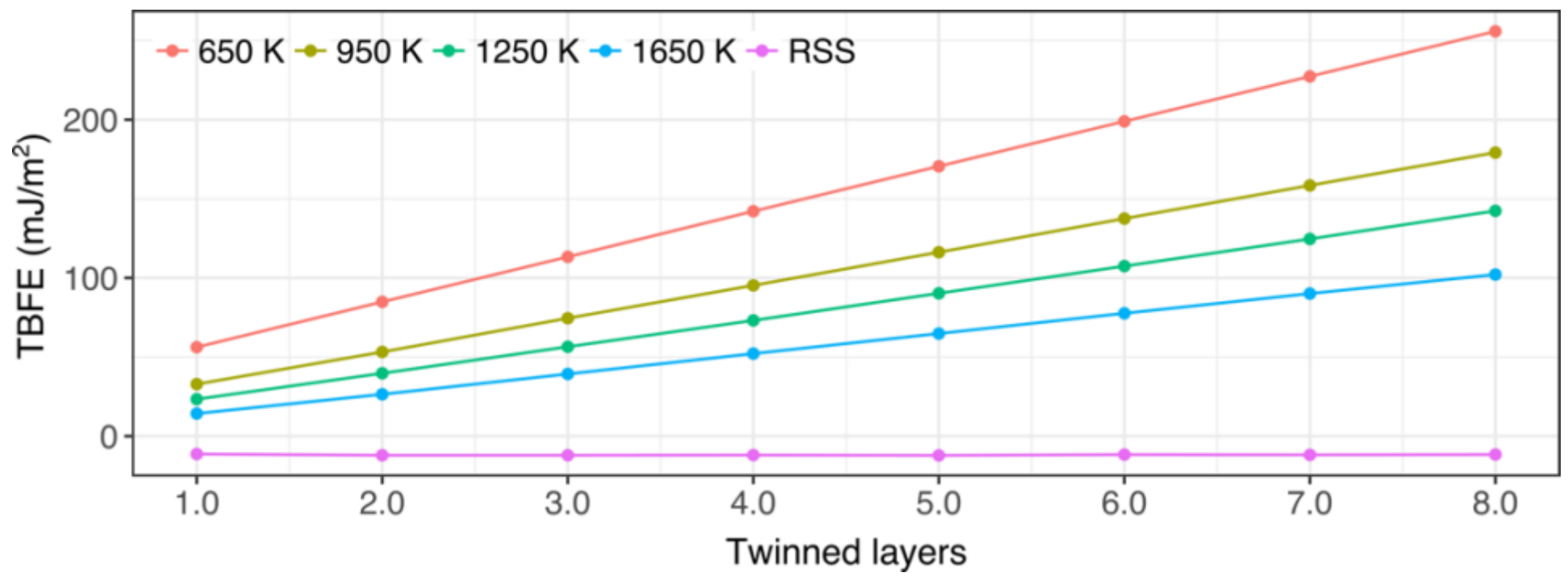

**Fig. S10 |** Variations of twin boundary fault energy with respect to twinned layers. Temperatures in legends are the annealing temperature. All twin boundary energy calculations are performed at 0 K.

## S4. Effects of LCO on elastic properties, lattice constant and cohesive energy

For basic property assessments, the dimensions of the sample are 10 nm × 10 nm × 10 nm, along the [100], [010] and [001] directions, respectively. Periodic boundary conditions are applied in all directions. All samples were annealed at each annealing temperature using the hybrid MD/MC simulations until the LCO converges. Each property at a specific annealing temperature was calculated at zero temperature and averaged over ten configurations equally distributed over the last one million MD/MC steps. The calculated elastic properties include $C_{11}$, $C_{12}$, $C_{44}$, bulk modulus $B$ and Poisson's ratio ν. In Fig. S11, $C_{11}$ is the average value of $C_{11}$, $C_{22}$ and $C_{33}$ of the cubic sample. $C_{12}$ is the average value of $C_{12}$, $C_{13}$, and $C_{23}$ of the cubic sample. $C_{44}$ is the average value of $C_{44}$, $C_{55}$ and $C_{66}$ of the cubic sample. As seen in Fig. S11a, generally, the elastic properties slightly increase with increasing LCOs except for a critical annealing temperature range where significant chemical ordering develops. The fluctuations at the critical annealing temperature range might be due to the anisotropies developed in elastic constants, as a result of significant chemical ordering. For example, the tensor components $C_{11}$, $C_{22}$ and $C_{33}$ may no longer be equal in magnitude, i.e., some of them become larger while some of them become relatively smaller such that the average is even smaller than that of a sample with weaker LCOs. However, with further increasing LCOs, the magnitudes of all components continue to increase despite that the anisotropy still exists, resulting in increasing average values again. As seen in Fig. S11b, with increasing LCOs, the average lattice constant slightly increases while the cohesive energy gradually decreases, suggesting a trend toward more stable states.

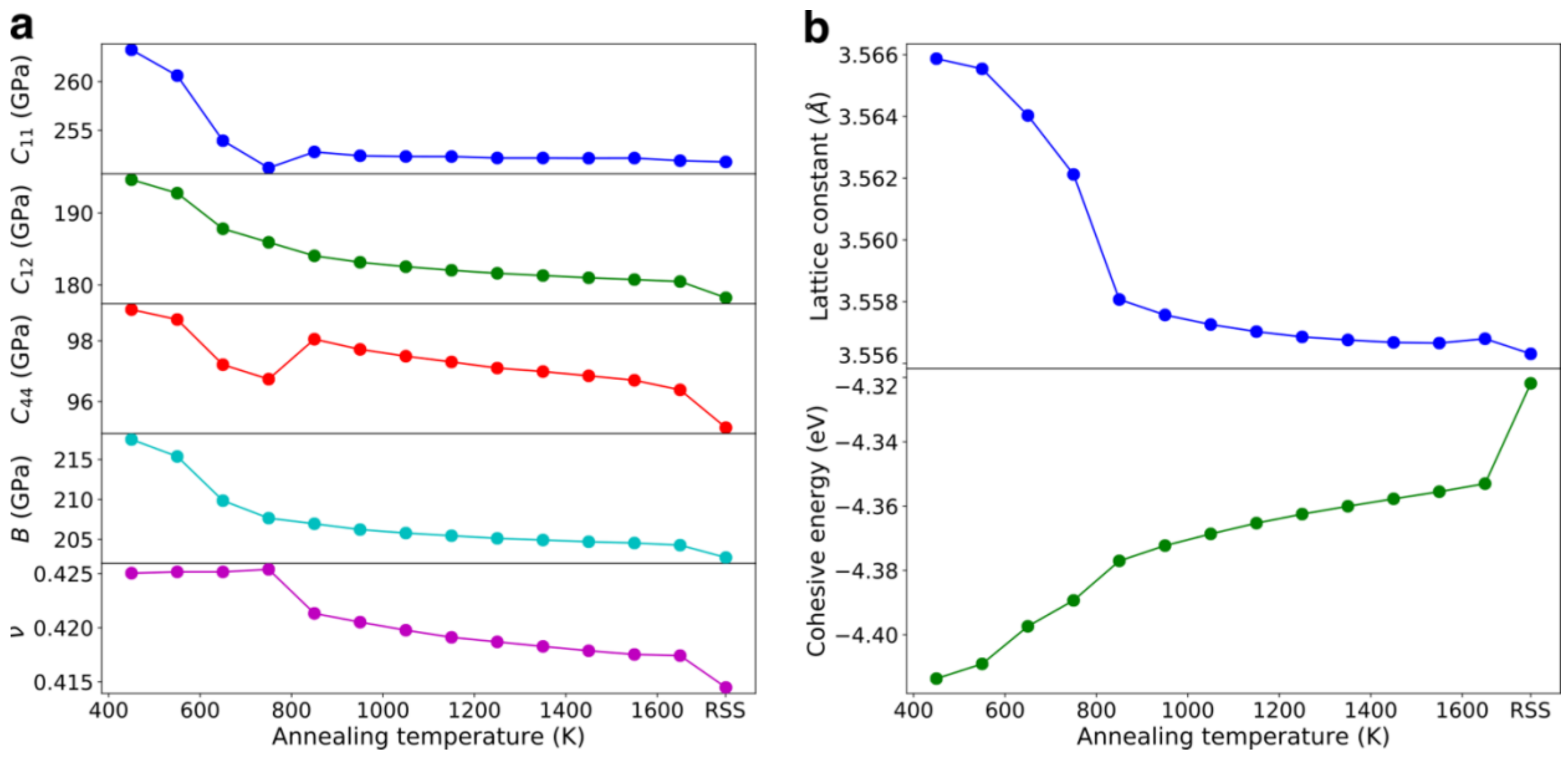

**Fig. S11 |** Annealing temperature dependent material properties. **a**, Elastic properties. **b,** Lattice constant and cohesive energy.

## S5. Nanoscale heterogeneity mapping for samples with relatively low annealing temperature

Fig. S12 shows the nanoscale property mapping, in a sample with $T_a$ = 650 K, to correlate local properties with LCOs. Fig. S12a shows the element distributions and the corresponding $\alpha_{ij}^{1}$ distributions. As expected, local composition fluctuations correspond to obvious spatial variations in $\alpha_{ij}^{1}$. However, no strong correlations are seen between these $\alpha_{ij}^{1}$ distributions and the distributions of local CSFEs (Fig. S12b, leftmost panel) and local APBEs (Fig. S12c, leftmost panel). Instead, the *change* of certain $\alpha_{ij}^{1}$ after introducing the fault, i.e., $\Delta\alpha_{ij}^{1} = \alpha_{ij}^{1,\text{fault}} - \alpha_{ij}^{1,\text{perfect}}$, correlates well with the distributions of both local CSFEs and local APBEs.

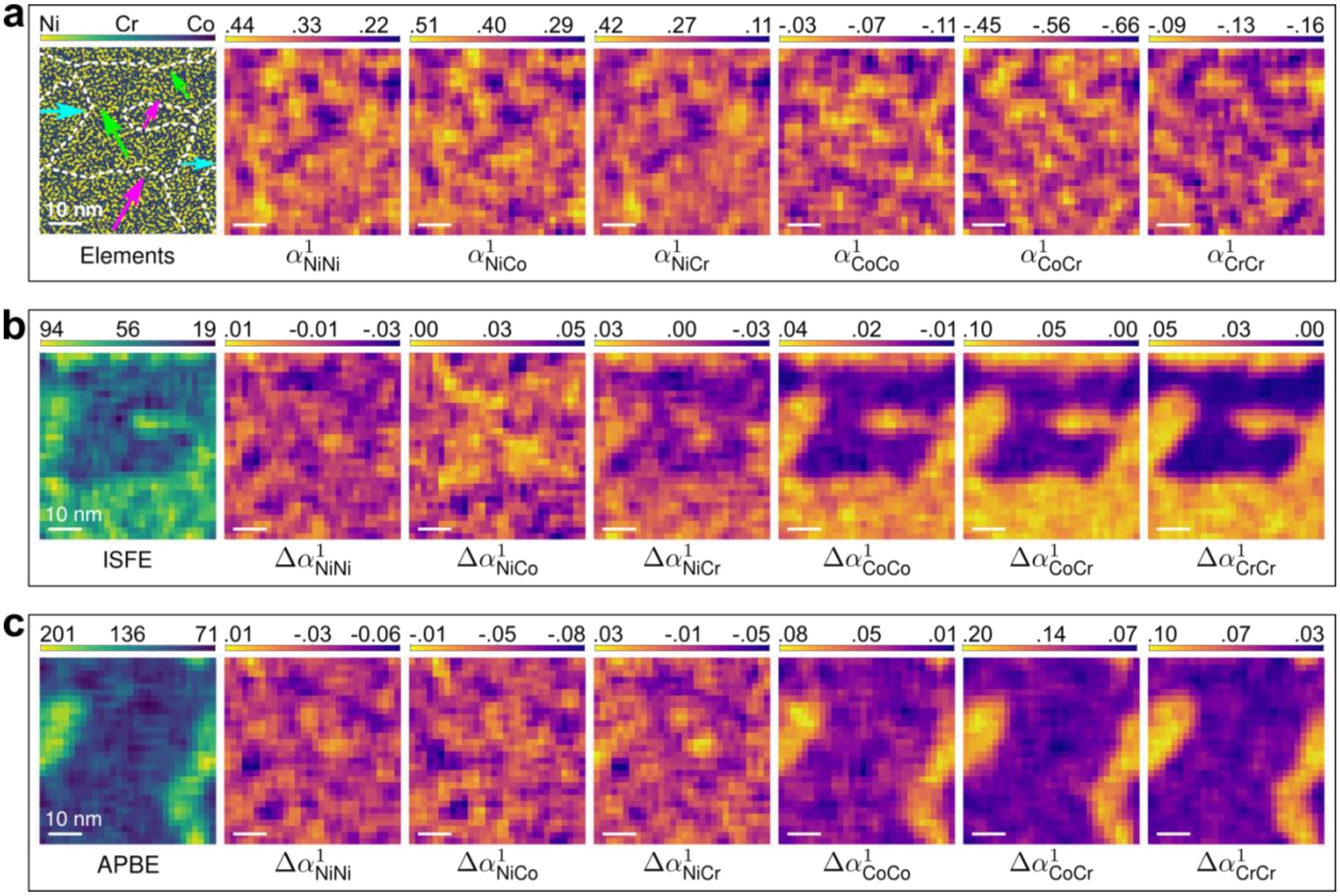

**Fig. S12** | Nano-scale heterogeneities and their correlations. The LCO in this sample was developed at $T_a$ = 650 K. **a**, Element distribution and the spatially varying local chemical order parameter. Arrows with different colors in the leftmost figure denote the orientations of the Co-Cr domains. Dashed lines are domain boundaries. **b**, Spatial distribution of local complex stacking fault energies and the corresponding changes of local chemical orders ($\Delta\alpha_{ij}^{1}$) due to the gliding of a leading partial dislocation. **c**, Spatial distribution of local anti-phase boundary energies and the corresponding changes of local chemical orders ($\Delta\alpha_{ij}^{1}$) caused by the gliding of a full dislocation. $\Delta\alpha^{1} = \Delta\alpha^{1,\mathrm{ISF}} - \Delta\alpha^{1,\mathrm{perfect}}$ for (**b**) and $\Delta\alpha^{1} = \Delta\alpha^{1,\mathrm{APB}} - \Delta\alpha^{1,\mathrm{perfect}}$ for (**c**). The value of each local area is calculated based on a fault area of 3.2 $nm^2$ and averaged over its 1st and 2nd nearest neighbor areas (9 areas in total including the central area). The scale bar is 10 nm.

Specifically, regions with smaller $\Delta\alpha_{\mathrm{CoCo}}^{1}$, $\Delta\alpha_{\mathrm{CoCr}}^{1}$ and $\Delta\alpha_{\mathrm{CrCr}}^{1}$ show lower CSFEs and APBEs, and vice versa (Fig. S12b-c). Furthermore, the local CSFE (Fig. S12b) and local APBE (Fig. S12c) distributions coincide with certain Co-Cr domains shown in Fig. S12a. For example, the relatively dark area in local CSFE mapping (Fig. S12b) corresponds to the green-arrow domain in Fig. S12a, while the darker region in local APBE mapping (Fig. S12c) matches the area occupied by the magenta and green-arrow domains in Fig. S12a. This suggests that the local fault energies are highly sensitive to the orientations of Co-Cr domains, in addition to the degree of Co-Cr LCO. For samples with much higher $T_a$, the orientation sensitivity would decrease as the Co-Cr clusters are too randomly oriented to form well-defined domains; the fault energy is more closely related to the degree of Co-Cr chemical order. The local fault energies only show weak correlations to Ni-related chemical order changes (Fig. S12b-c). This is because FCC Ni precipitates are isotropic with respect to the <112> shear. Thus each of them may experience similar chemical order changes after shearing, and the whole region looks more uniform than non-Ni chemical order changes (i.e., $\Delta\alpha^{1}_{\mathrm{CoCo}}$, $\Delta\alpha^{1}_{\mathrm{CoCr}}$ and $\Delta\alpha^{1}_{\mathrm{CrCr}}$).

## S6. Long dislocation line gliding in copper

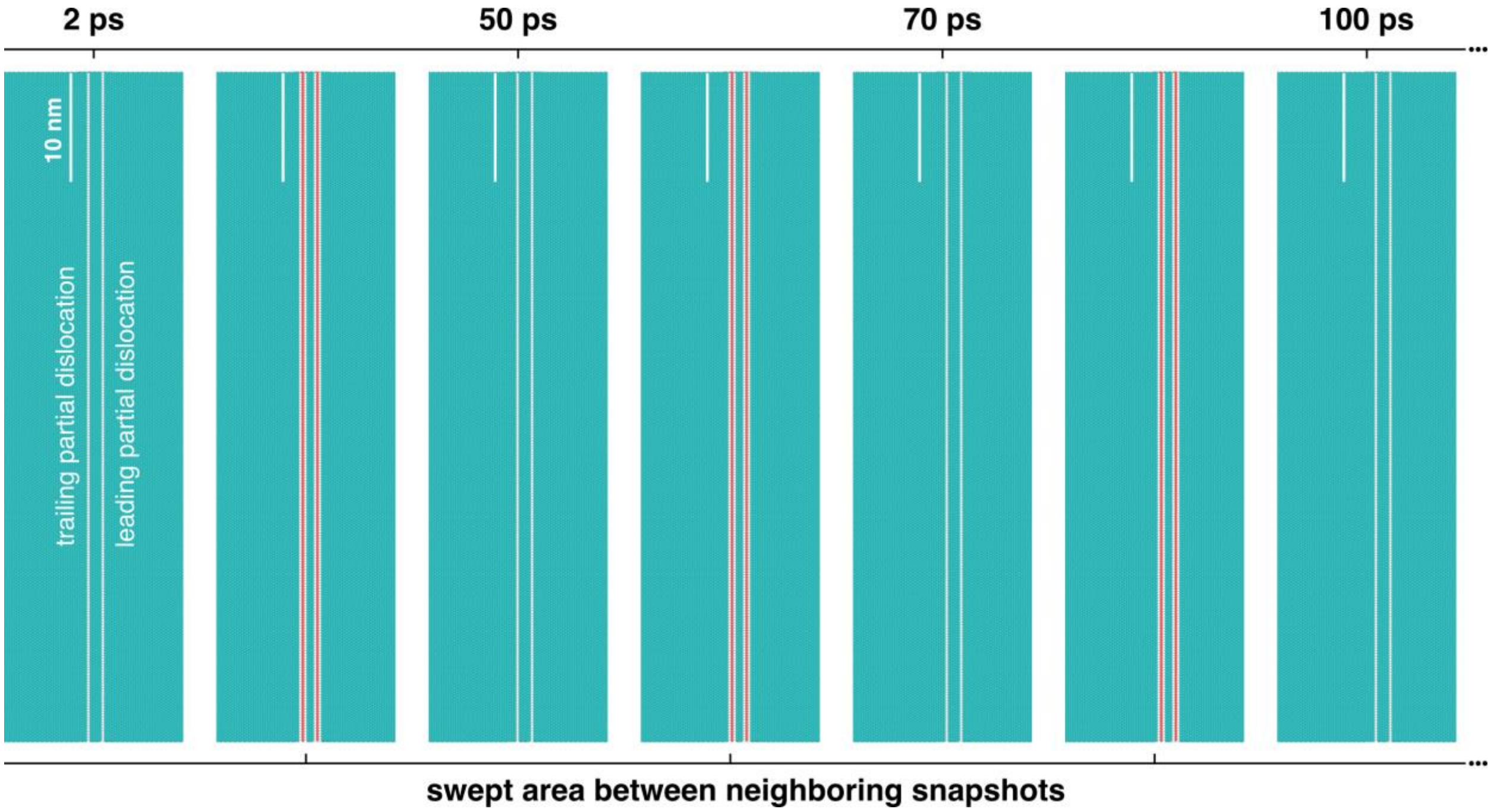

**Fig. S13 |** Uniform and smooth glide of extended screw dislocation in Cu. The initial dislocation configuration is shown by the snapshot at 2 ps. This extended screw dislocation is then subjected to a constant shear stress of ~ 10 MPa at 300 K. The configurations at 50 ps, 70 ps and 100 ps demonstrate a forward glide process, with the uniform swept areas highlighted in red. All configurations are obtained by quenching the MD configurations to zero K and zero stress and energy minimization.

## S7. Sample size effects on Peierls barrier calculation

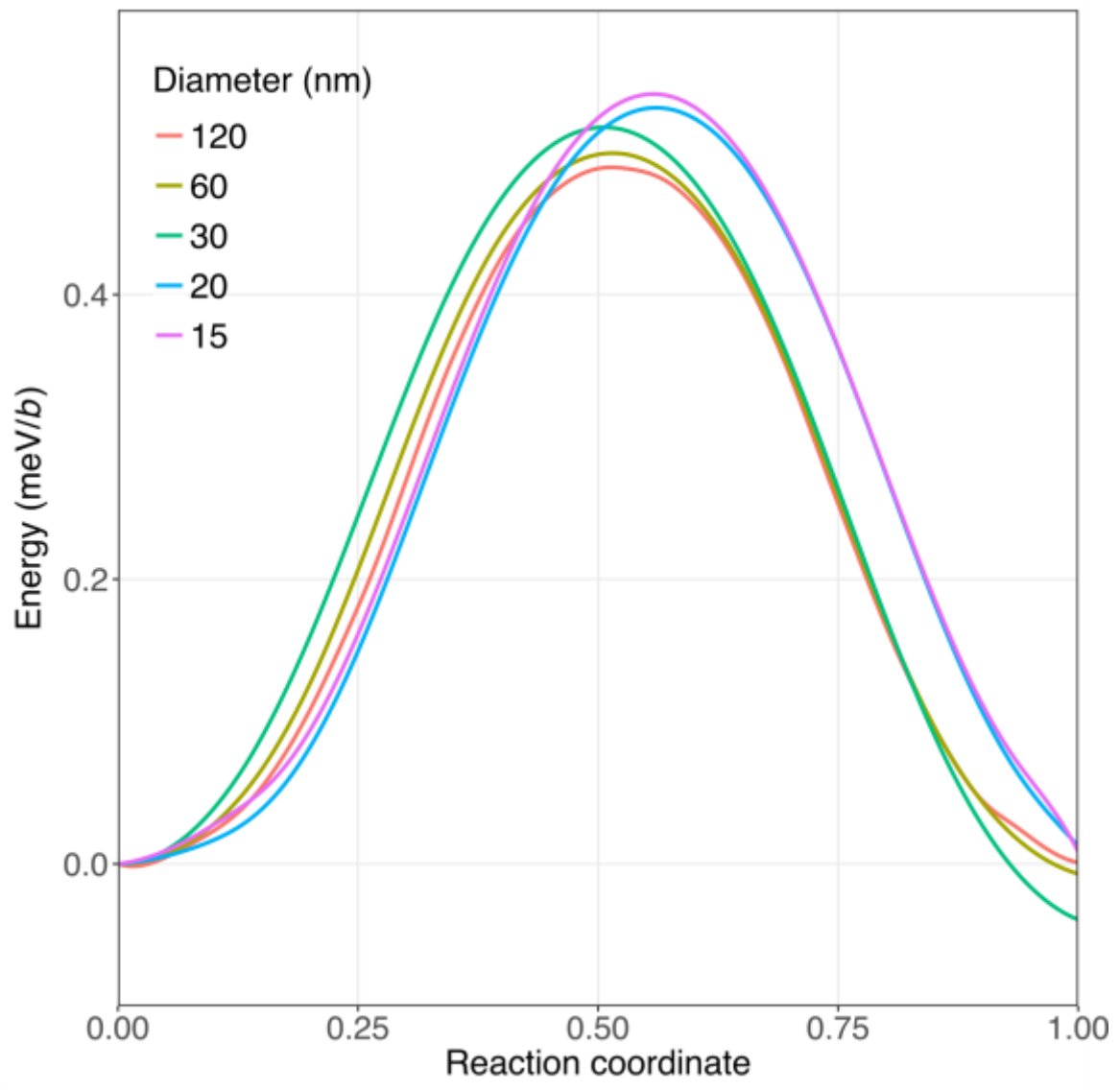

**Fig. S14** | Sample size (cylinder diameter) effect on the Peierls barrier. The calculations are based on cylinder configurations, using an EAM potential for Al[15]. All cylinder configurations have an axial length of 4*b*, where *b* is the magnitude of the full Burgers vector. A screw dislocation (with **b** along the axial direction) was introduced into the cylinder configuration following the anisotropic elasticity theory. Then atoms belonging to the outer layer with a thickness greater than two times the potential cutoff are fixed during the subsequent energy minimization. The free-end string method[16] was used for the minimum energy path calculations. As seen, with smaller diameters, the Peierls barrier only slightly increases, showing a limited size effect on the barriers. Likewise, since we used the same size for all the samples in our main text, the sample size should have a negligible effect on the relative magnitude of the Peierls barriers evaluated for samples with different $T_a$.

## S8. Effective activation barriers for nanoscale segment detrapping events

We carried out the minimum energy path (MEP) calculations in samples under different processing conditions including RSS, $T_a$ = 1350 K, $T_a$ = 950 K and $T_a$ = 650 K. For each processing condition, we used 30 different samples under different stress levels to collect sufficient data points for statistical comparison. Fig. S15 shows the effective activation barriers for all types of samples. From the plot, several important observations can be made. First, at a specific local stress level, the activation barrier is not a single value; instead the magnitude of the activation barrier spans a wide range. Such a spread of activation barriers results from the nanoscale heterogeneities as shown in Fig. 3 in the main text. Second, for a given stress level, increasing LCO (RSS → $T_a$ = 1350 K → $T_a$ = 950 K → $T_a$ = 650 K) leads to an increasingly wider range of barrier heights, suggesting additionally higher barriers in the more ordered samples. Third, for samples processed at each $T_a$, the barrier range gradually narrows with increasing local shear stress, approaching a vanishing barrier. Such barrier-range expansion and athermal stress limit increase with increasing LCOs suggest remarkable LCO-induced strengthening as discussed in the main text.

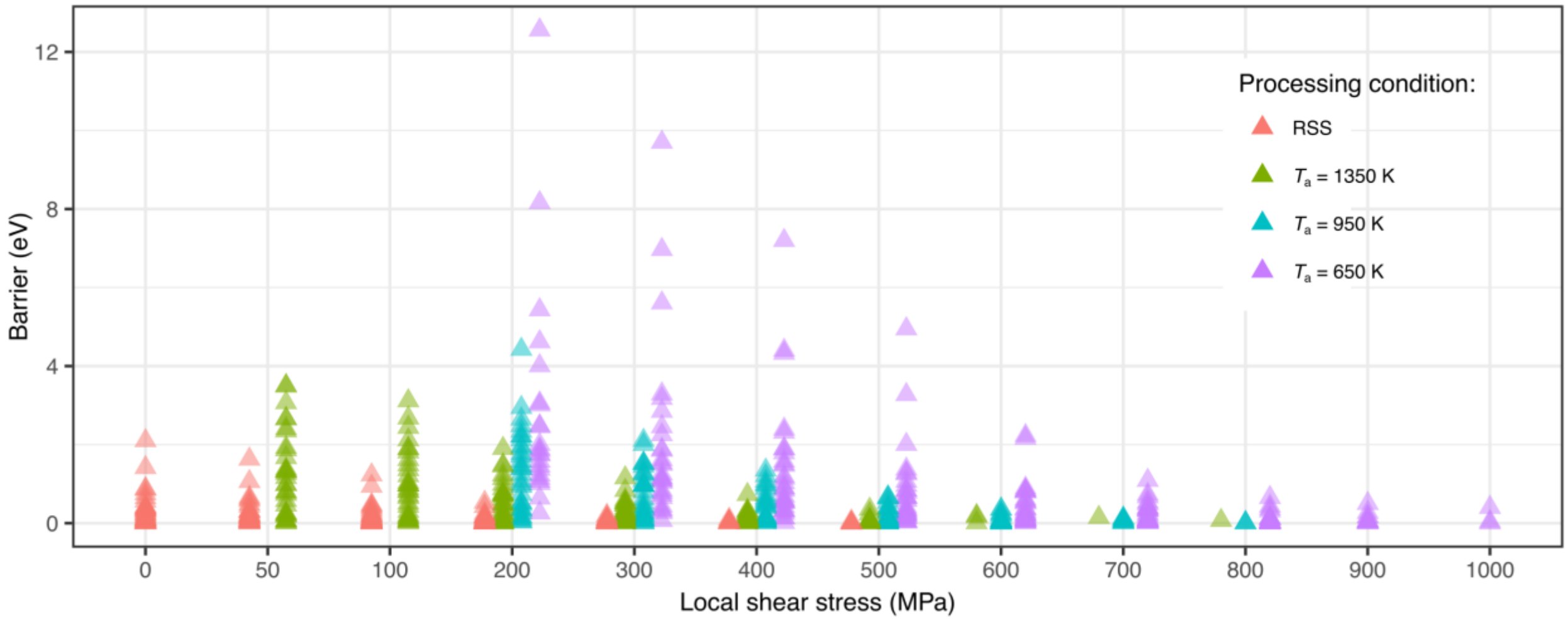

**Fig. S15** | Effective barriers calculated for nanoscale depinning processes in different samples.

## S9. Activation free energy and activation (shear) stress

The activation volumes at finite temperatures can be obtained from the stress dependence of the activation free energies, $\Omega(\tau,T) = -\partial Q(\tau,T)/\partial\tau$. Following previous work[3,17,18], the activation free energy is approximated as $Q(\tau,T) = Q_0\left(1-\frac{T}{T^*}\right)\left(1-\frac{\tau}{\tau_{\mathrm{ath}}}\right)^{\alpha}$, where $Q_0$ is the activation energy at zero-temperature and zero-stress, $\tau_{\mathrm{ath}}$ is the athermal strength, $T^*$ is a characteristic temperature for estimating the activation entropy and $\alpha$ is a constant controlling the stress dependence of $Q$. The characteristic temperature $T^*$is evaluated as $\frac{1}{T^*} = \frac{1}{\mu_0}\frac{\mathrm{d}\mu}{\mathrm{d}T}$, where $\mu_0$is the zero-temperature shear modulus and $\frac{\mathrm{d}\mu}{\mathrm{d}T}$ is the changing rate of shear modulus with respect to temperature. For each type of sample, we calculated the shear modulus at 0 K, 50 K, 100 K, 150 K, 200 K, 250 K and 300K, respectively. At each temperature, we carried out ten calculations to obtain the statistical mean shear modulus. Then $\frac{\mathrm{d}\mu}{\mathrm{d}T}$ can be obtained by considering all the temperature-dependent shear modulus. Finally, the $T^*$for different samples were obtained as follows: 6909 K, 7653 K, 8973 K and 8446 K, for samples of RSS, $T_a$ = 1350 K, $T_a$ = 950 K, $T_a$ = 650 K, respectively. Note that $T^*$for samples with $T_a$ = 650 K is slightly lower than that for samples with $T_a$ = 950 K. This is consistent with the results in Fig. S11, where a temporary shear modulus softening shows up during the significant local chemical ordering process around $T_a$ = 850 K. The reason for this shear modulus softening was explained in section S4.

The intrinsic yielding strength can then be evaluated by solving Orowan's equation

$$\dot{\gamma} = \rho_{\mathrm{m}} b \bar{v} = \rho_{\mathrm{m}} b v_0 \bar{d} \exp\left(-\frac{Q(\tau,T)}{k_{\mathrm{B}}T}\right) \tag{1}$$

where $\rho_{\mathrm{m}}$ is the mobile dislocation density (~$10^8$ m$^{-2}$, assuming one dislocation in a grain with typical grain size of 100 μm), $\bar{v}$ is the average dislocation velocity which can be further expressed in the form of $\bar{v} = v_0 \bar{d} \exp\left(-\frac{Q(\tau,T)}{k_{\mathrm{B}}T}\right)$, with $v_0$ the attempt frequency (~$10^{12}$ s$^{-1}$, see Section 11), $\bar{d}$ the average distance a dislocation segment moves through an activation event (~1.1 nm, an average value over 50 events in different samples under different stresses), and $k_{\mathrm{B}}$ the Boltzmann's constant.

## S10. MD simulations of extended edge dislocations

Edge dislocations were simulated using the equilibrium configurations from the hybrid MD and MC simulations. The as-prepared sample was first replicated along *y*[111] direction by 2 times and then replicated along $z[1\bar{1}0]$ direction by 2 times. As schematically illustrated in Fig. S16, the simulation box has a geometry of $X[11\bar{2}]$ 52 nm × $Y[111]$12.5 nm × $Z[1\bar{1}0]$ 111 nm. An edge dislocation was introduced according to the following procedure. First, we centered the desired slip plane to the center of the simulation box and turned off the periodic boundary conditions along *y* direction. Then we divided the simulation box into upper and lower slabs. For the lower slab, an atomic plane perpendicular to the $z[1\bar{1}0]$ direction was deleted. In this way, an extra plane was introduced in the upper slab. Atoms beside the deleted plane were rescaled to fill up the extra space. Then the simulation box was uniformly scaled such that the *z* direction length was reduced by half of the Burgers vector. Energy minimization was further performed to relax the defect configuration, after which the as-introduced dislocation configuration was relaxed at 300 K for 100 ps and then the temperature was increased to 1000 K to relax for an extra100 ps. The high-temperature relaxed configuration was then cooled down to 300 K and the boundary condition along the dislocation motion direction was also changed to free surface. Finally, after a short period of relaxation at 300 K, a constant shear strain rate of $1\times10^7$ s$^{-1}$ was applied by assigning a constant velocity in *z* direction to the top surface atoms (in *y* direction) while the bottom surface atoms were fixed. The introduced dislocation is ~52 nm long to allow sufficient variations of core configurations along the line sense direction. The relaxed simulation configuration is schematically illustrated in Fig. S16.

As shown in Fig. S17, the full dislocation generally dissociates into partial dislocations with variable curvatures along dislocation line sense direction. The dissociation widths in samples of higher $T_a$s are generally larger than those of relatively lower $T_a$s. For random solid solutions, the separation between partial dislocations becomes significantly large after relaxation at 1000 K, which is no longer appropriate for the subsequent shear deformation. Instead, the configuration shown in the left panel of Fig. S17a was used to perform the shear deformation. The dissociation width in samples of $T_a$ = 1650 K remains largely constant before and after relaxation at 1000 K. The dissociation width in samples of both $T_a$ = 1350 K and $T_a$ = 950 K becomes smaller after relaxation at 1000 K. These results demonstrate that both the lattice resistance and the CSFE affect the dislocation dissociation width. For samples of $T_a$ = 650 K and $T_a$ = 350 K, due to the large APBE and CSFE, the as-introduced dislocation is already able to overcome the lattice resistance to form relatively narrow dislocation cores which largely remain constant after relaxation at 1000 K. Overall, the dislocation core configurations, in terms of the variation along the dislocation line and dissociation behavior, are strongly dependent on $T_a$ and thus local chemical ordering.

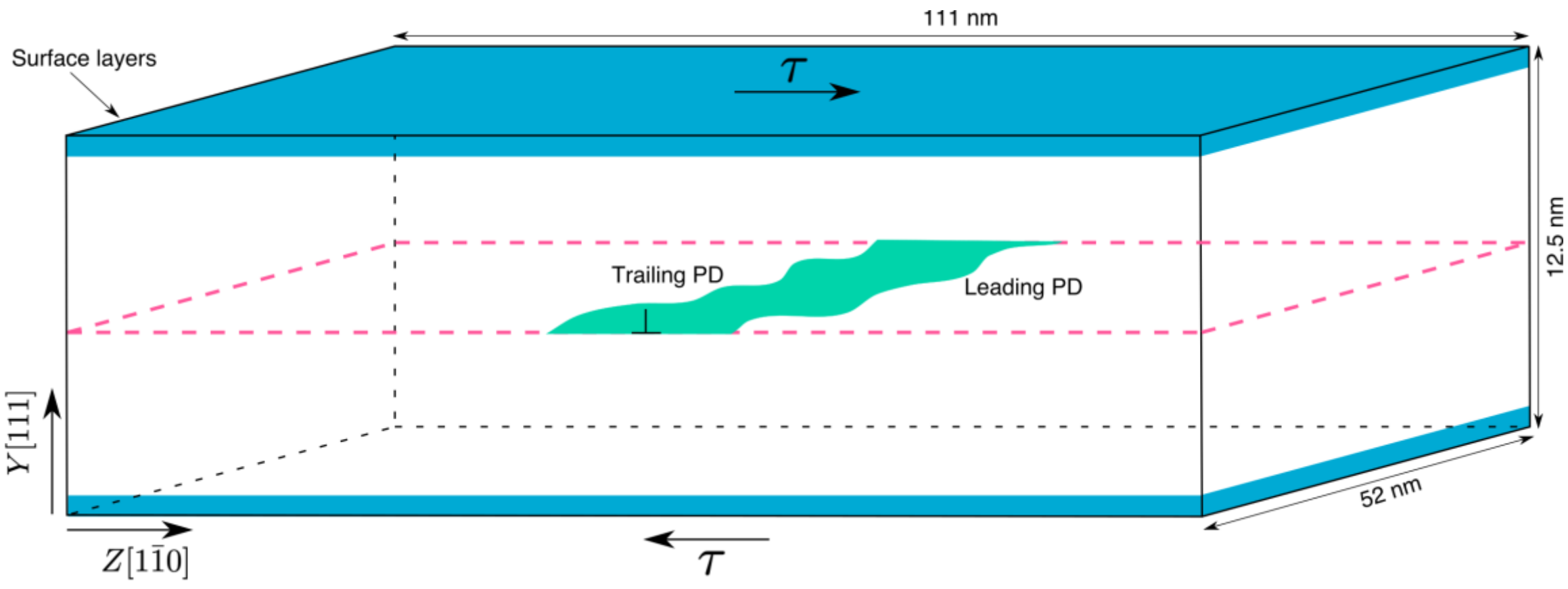

**Fig. S16** | Schematic illustration of the simulation setup for edge dislocations. Periodic boundary conditions are applied in dislocation line direction (i.e., *X* direction), while free surfaces are used in both *Y* and *Z* directions. The introduced edge dislocation generally dissociates into two partial dislocations (PD) with wavy dislocation lines. To apply a constant strain rate, a constant velocity in the *Z* direction was assigned to the top surface atoms (region colored in blue) while the bottom surface atoms were fixed.

Fig. S18 shows the shear stress vs. time curve (middle panel) and the corresponding snapshots of dislocations in samples with $T_a$ = 650 K (upper panel) and $T_a$ = 1350 K (lower panel), respectively. A significant stress drop on the curve corresponds to the movement of a partial dislocation. Several interesting phenomena arise from the shear responses of these dislocations.

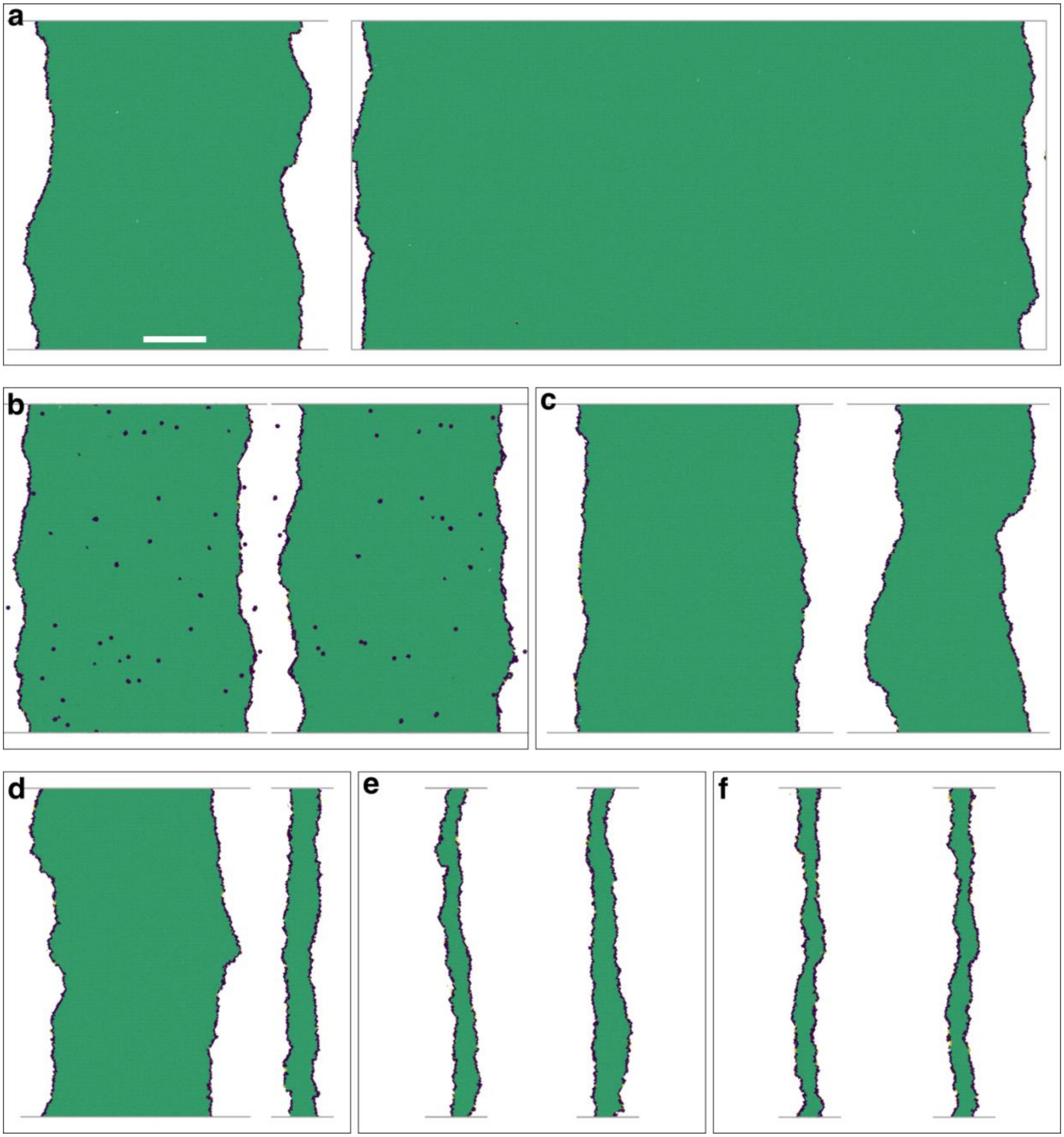

**Fig. S17** | Dislocation core configurations for random sample (**a**), sample annealed at 1650 K (**b**), sample annealed at 1350 K (**c**), sample annealed at 950 K (**d**), sample annealed at 650 K (**e**) and sample annealed at 350 K (**f**). In each figure, the left configuration is the as-introduced dislocation core after relaxation at 300 K for 100 ps and the right configuration is after relaxation at 1000 K for 100 ps and then cooled down to 300 K. The scale bar is 10 nm. Atoms in stacking fault are rendered in green while atoms in dislocation core or point defects are in purple. The small atom clusters in (**b**) are formed by thermally induced point defects due to annealing at temperatures not far from the melting point.

First, as $T_a$ decreases, a significant increase in the resistance to both the leading partial dislocation and the trailing partial dislocation is observed. Specifically, for samples of random solid solution, $T_a$ = 1350 K, 950 K and 650 K, the critical shear stresses for the leading partial dislocation are 0.35 GPa, 0.52 GPa, 0.64 GPa and 1.10 GPa, respectively. The strengthening is also seen in the increasing critical shear stress to drive the trailing partial dislocation, 0.60 GPa, 0.80 GPa, 1.00 GPa and 0.92 GPa, respectively. Such pronounced strengthening effects can be attributed to the increasing LCOs with decreasing $T_a$: it costs more energy for a partial dislocation to break loose from the increasingly stronger LCOs as $T_a$ decreases, consistent with the results on CSFE and APBE in the main text.

Second, the atomic scale LCO and the resultant nanoscale segregations/clusters act as strengthening obstacles to dislocation motion. Indeed, dislocation lines frequently show remarkable local curvatures during motion (see Movie S1 and Movie S2 for dynamic processes, for $T_a$= 1350 K and $T_a$ = 650 K), which indicates strong impedance of local dislocation segments that would otherwise move rather smoothly. Such a rugged advancing dislocation line is consistent with the abundant nanoscale heterogeneities shown in both Fig. 1 and Fig. 3 in the main text, much like sailing in the choppy "sea" of LCO. The extra resistance leads to an elevated Peierls stress. The heterogeneous dislocation motion may create significant line tension modifying the magnitude of the local stress. In this case, it is not sufficient to use the externally applied stress to specify the stress-sensitive local activation barriers; the local stress dependences of the activation barriers shown in Fig. 5a are thus more intrinsic.

Third, the trailing partial dislocation is much harder to move than the leading partial dislocation. For example, for the random solution and the $T_a \geq 950$ K samples, the critical shear stress to drive a trailing partial dislocation is ~60% higher than that to drive a leading partial dislocation. Such higher stresses to drive trailing partial dislocation is not due to our simulation setup. Specifically, the current simulation box has a large in-plane size (Fig. S16) of 52 nm (dislocation line direction) × 111 nm (dislocation motion direction), so the box size effects should have been minimal. Instead, the higher stresses needed to drive the trailing partial dislocations are caused by the relatively larger energy cost to eliminate the complex stacking fault (CSF) and create local antiphase boundaries (APB). This can be seen from differences between APBE and CSFE, i.e., APBE – CSFE is a larger value when compared to CSFE. For example, for $T_a$ = 1350 K and $T_a$ = 950 K, the differences between APBE and CSFE are 52.24 mJ/m$^2$ and 58.37 mJ/m$^2$, respectively, while their CSFEs are only 9.11 mJ/m$^2$ and 22.58 mJ/m$^2$, respectively. Thus, the average energy penalty to eliminate CSF is considerably higher than that to create CSF, resulting in higher stresses to drive trailing partial dislocations. In contrast, for samples with $T_a$ = 650 K, APBE – CSFE is 60.60 mJ/m$^2$ while the CSFE is 60.69 mJ/m$^2$, thus the stresses needed to drive leading partial dislocation and trailing partial dislocation should be comparable to each other. However, for samples with such lower $T_a$, the stresses to drive leading and trailing partial dislocation would also depend on the local Co-Cr domain orientations (see Fig. S12).

The more energetically favorable motion of the leading partial dislocation should significantly enhance the formation of uniformly distributed SFs and a high population of very thin (nano-)twins. Indeed, profuse SFs and nanoscale twins were reported to be responsible for the strong work hardening and good ductility of the HEAs/MEAs[19–21,2,22]. Experiments also reported the preference for planar slip in HEAs[19,20,23–29], which would be expected from repeated operation of dislocation that eliminates the LCO on some specific planes (see Fig. 2 in main text). In other words, the plethora of experimental observations in HEAs/MEAs, including the local SFE, faults and nanotwins versus extended dislocations, planar slip, and the very different strength and hardening behavior for HEA/MEA samples processed at different annealing or homogenization temperatures, can now all be explained under the same umbrella of the variable LCO.

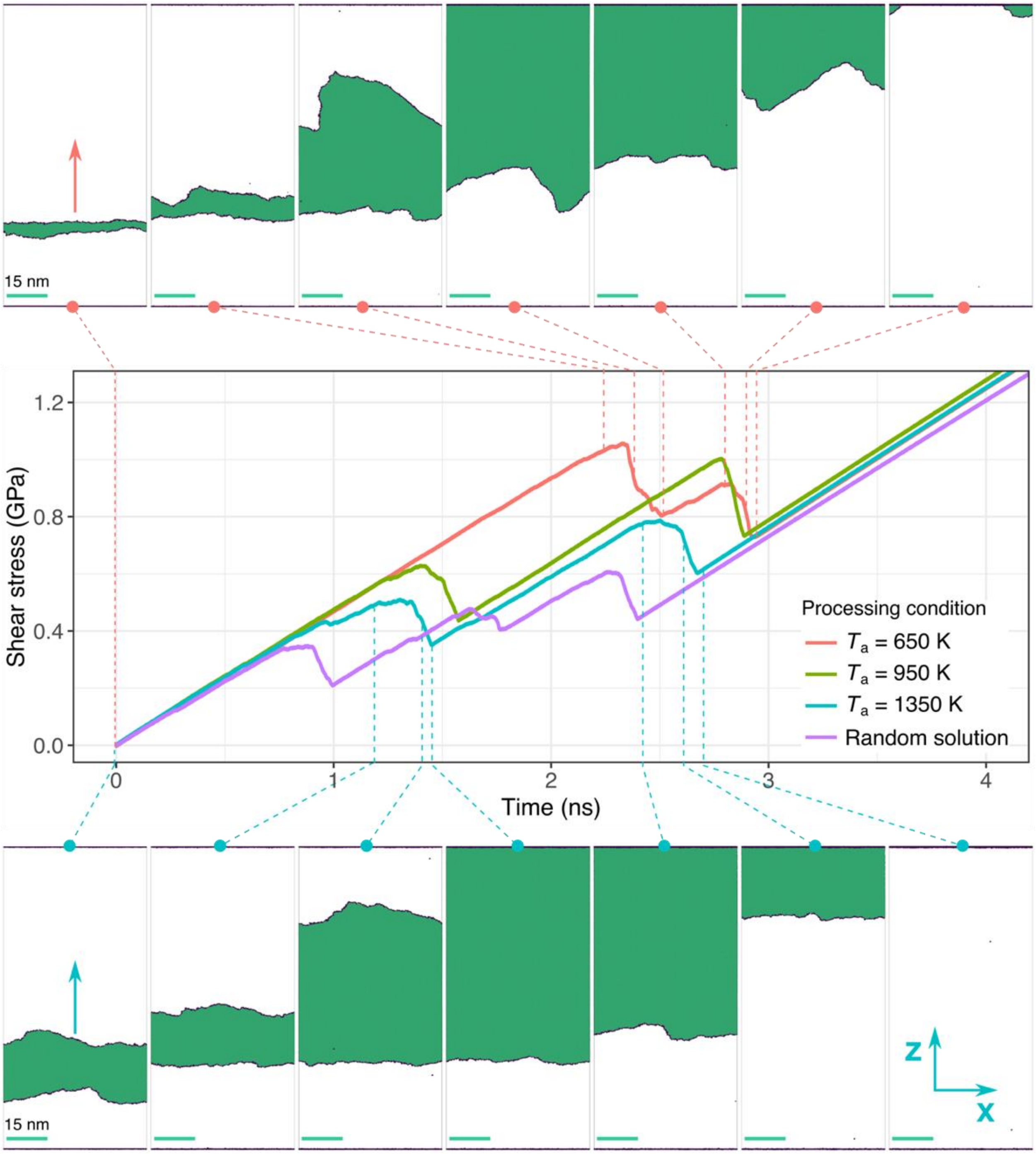

**Fig. S18 |** Local chemical order induces strengthening. Dislocations in the random solution sample and samples annealed at 1350 K, 950 K and 650 K are subjected to simple shear at 300 K with a constant shear strain rate of $1\times10^7$ $s^{-1}$. The middle panel shows the shear stress vs. time curve, showing the strengthening due to increasing LCO that triples the stress needed to move the dislocation. The upper and lower panels show snapshots of dislocation configurations in samples annealed at 650 K and 1350 K, respectively. The arrows indicate the motion direction of the dislocation. $X$ is along the $[11\bar{2}]$ direction and $Z$ is along the $[1\bar{1}0]$ direction. The scale bar in each snapshot is 15 nm.

## S11. Verification of string method with NEB calculations

For minimum energy path calculations, we followed the suggestions made by Nöhring and Curtin[30] that the string method could be more robust for complex concentrated alloys. However, we did compare the string method results with NEB calculations. Fig. S19 shows an example of the MEPs calculated by string method and NEB method, respectively. As can be seen, the overall profiles of both MEPs are very similar to each other; those small differences should be due to the specific constraints used in updating intermediate replicas (e.g., numerical reparameterization for string method vs. nudged potential forces for NEB method). Despite these slight differences, the major saddle points from both methods coincide with each other very well, suggesting very good consistency. We also extensively tested the convergence criteria used in the string method. Overall, results obtained using the current criteria are comparable to well converged NEB calculations. For example, in Fig. S19, the NEB calculation was considered as converged only when the forces on each replica are less than 0.001 eV/Å.

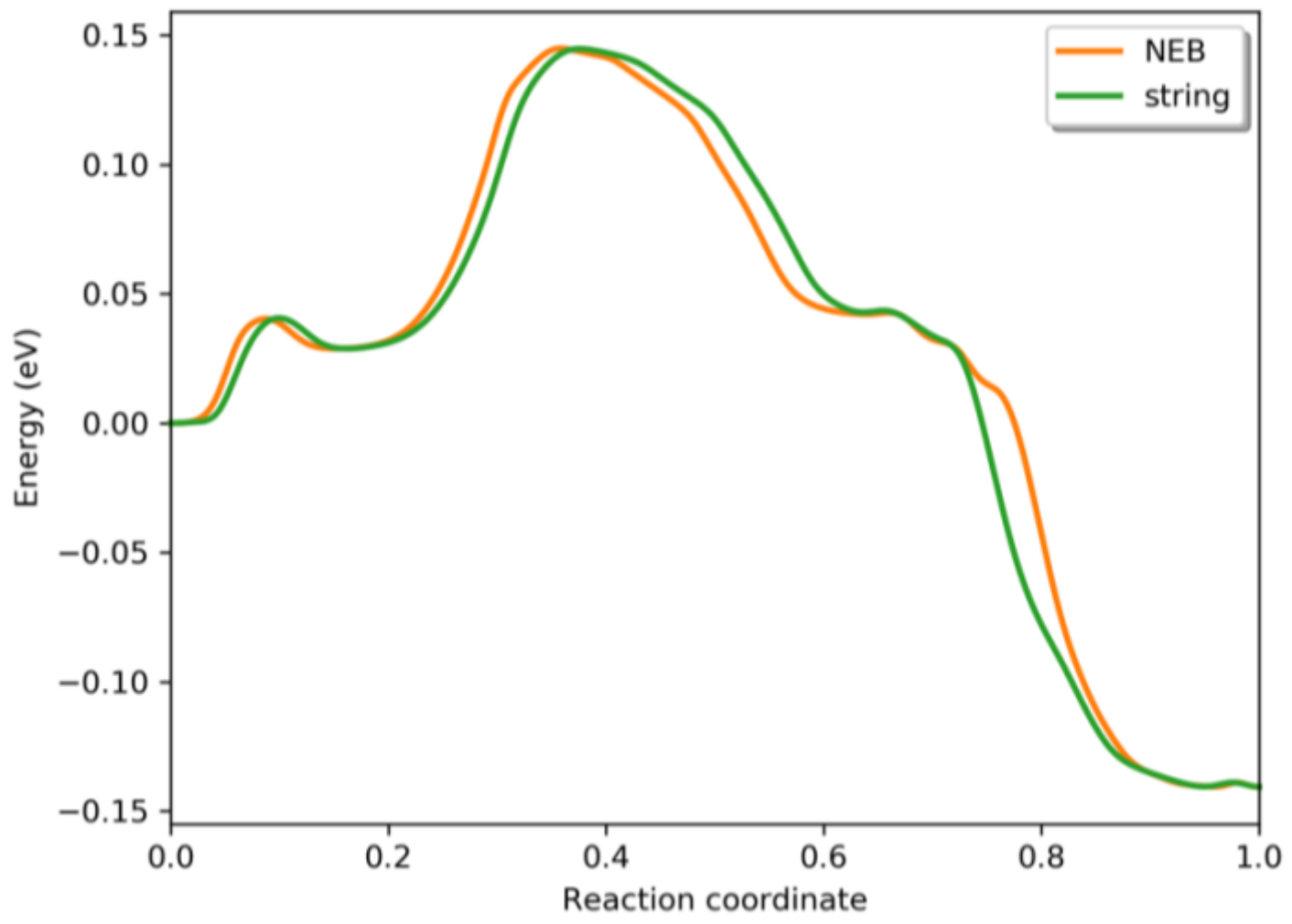

**Fig. S19 |** Comparison between the minimum energy paths calculated using the string method and the NEB method. This example corresponds to Fig. 4b in the main text.

It should be noted that for complex rugged energy landscape, there may be multiple transition pathways between the same set of initial/final states. The calculated MEP from either string method or NEB method is only one possible transition path; more advanced methods (e.g., finite-temperature string method and forward flux sampling etc.) are required to obtain the accurate transition rate of a *specific event*. However, in this work, we are not focusing on a single event but rather on sampling possible nanoscale segment detrapping events in the entire sample. For this purpose of the current work, string method calculations/NEB calculations over many randomly chosen events would draw a sample set that reflects the characteristics of the entire distribution. Such calculated barriers can be further verified using Arrhenius plot. For example, Fig. S20 plots the natural logarithm of partial dislocation velocities ($\ln(v)$) against $1/(k_B T)$ in RSS samples subjected to a shear stress of 200 MPa. As can be seen, the fitted average activation energy in this case is 0.052 eV which is close to the string-method calculated activation energy 0.085 eV (with a standard deviation of 0.0625 eV). On the other hand, the fitted intercept $\ln(A)$, where $A = v_0 \bar{d} \exp(\Delta S)$, is 7.8. As we already know $\bar{d}$ and $\exp(\Delta S)$ (from the thermodynamic compensation rule in Section S9), the attempt frequency is estimated to be $2.1 \times 10^{12} \mathrm{s}^{-1}$, which is in the expected range from $10^{11} \mathrm{s}^{-1}$ to $10^{13} \mathrm{s}^{-1}$.

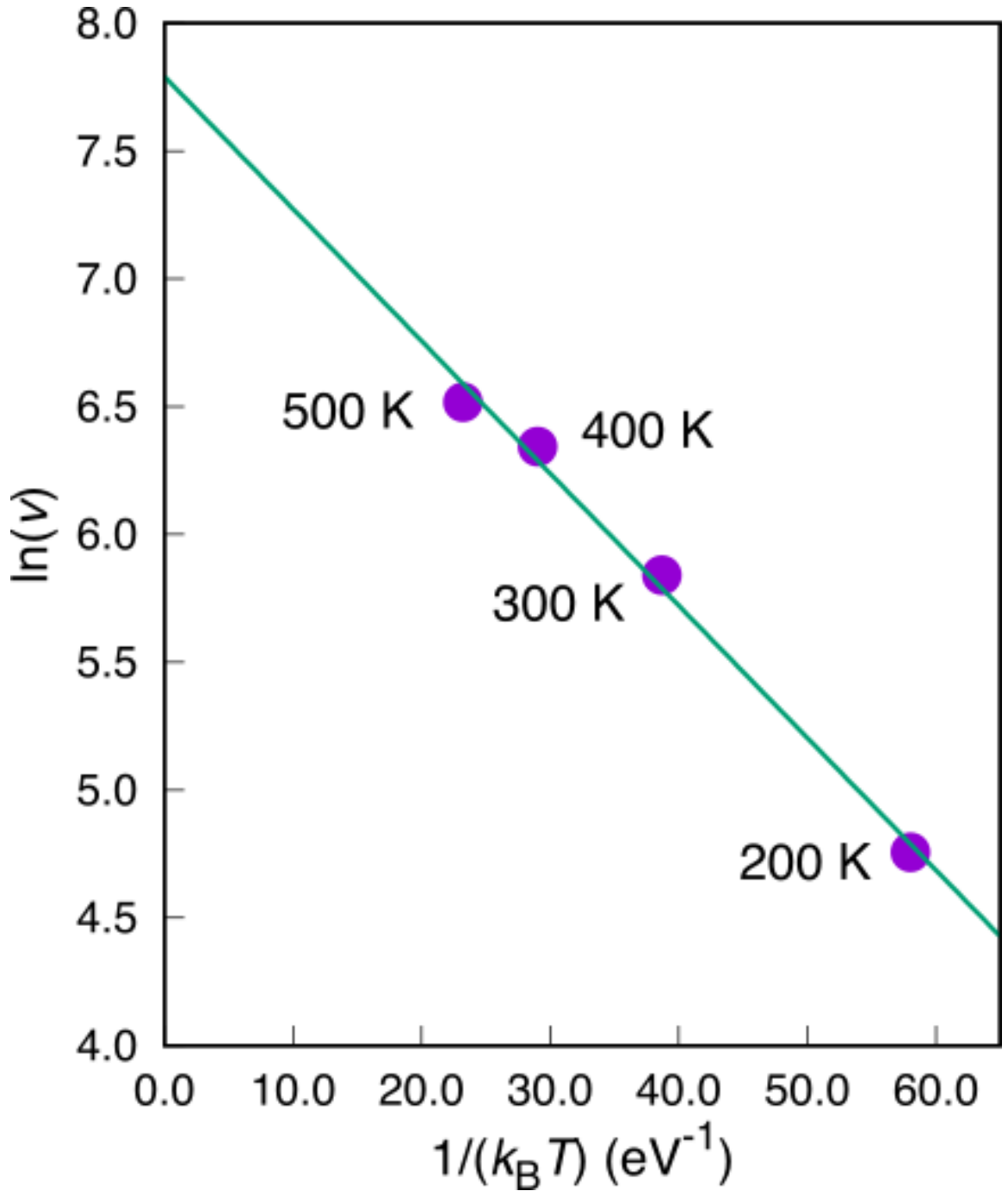

**Fig. S20 |** Arrhenius plot of partial dislocation velocities ($\ln(v)$) vs. $1/(k_B T)$ in RSS. The applied shear stress is 200 MPa. The dislocation line length is ~30 nm and the glide distance is ~64 nm. Each data point was averaged over 5 different RSS samples. The Arrhenius-plot-informed average activation energy is 0.052 eV (the negative slope) which is close to the string-method calculated activation energy 0.085 eV with a standard deviation of 0.0625 eV.

## S12. Discussion on the magnetic effects

There have been no well-accepted potentials so far for multi-component HEAs (the published MD simulation papers had to use some "average atom" potentials), seriously hampering atomistic modeling of these new concentrated alloys. As detailed in Section S1, our potential was developed based on non-magnetic DFT calculations and experimental inputs. Spin polarization was not explicitly considered. Thus, our potential is an "empirical potential" to model the Ni-Co-Cr system and it is not intended for situations with significant magnetic effects. In other words, the as-developed EAM potential is a highly optimized atomistic model for equi-atomic multicomponent system, the best there is for HEAs; however, it is not "realistic" to the point that it can describe magnetic transition and spin polarization effects.

In fact, atomistic models employing classic potential formalisms (e.g., the embedded-atom-method adopted here) never fully account for the electron spins. This is because of i) the limitation of the potential formalism: the goal is the construction of a potential that allows handling of large-scale simulations; but this is achieved at the expense of the accuracy of first-principles calculations. ii) The level of difficulty in accurately describing the magnetic states of the system. In practice, the magnetic effect is usually not explicit in such classic MD potentials and simulations. Even for pure Ni, with a Curie temperature of ~627K, MD simulations do not have the predictive power for its magnetic transition.

In general, atomistic models developed using the force-matching method will not be as accurate as DFT calculations. In our development of the Ni-Co-Cr potential, despite of our best effort, the average energy difference between the DFT data and the EAM potential is ~20 meV/atom (see Section S1). In the field of potential development, this margin of deviation is normally already considered a highly optimized interatomic potential for atomistic MD modeling. But the empirical potential could fail to capture the magnetic effects, if the resultant energy difference is only several meV/atom. This is essentially inherent to the empirical potential development, rather than our potential fitting procedure itself.

Then the next question is "how large are the magnetic effects in NiCoCr". We start our discussion from the random solution at this composition. Our spin-polarized DFT calculation shows a very small energy difference between the magnetic phase and the non-magnetic fcc phase, 2 meV/atom, and that between magnetic and non-magnetic hcp states is around 4 meV/atom (see Fig. S20 and its caption for methods). The SFE of the magnetic state is slightly more negative than the non-magnetic state by a difference of ~ 5-10 mJ/m2. Our present model can capture the energy difference between HCP and FCC NiCoCr random solid solutions. This is discussed in the context of intrinsic stacking fault energy in the main text. Our potential was optimized to reproduce the negative stacking fault energy of NiCoCr even without considering the magnetic effect. Both our calculations and Niu's work[4] clearly indicate that the HCP phase is favored over the fcc phase with or without magnetic contributions (see Fig. S20) in the equiatomic NiCoCr alloy. In the CrMnFeCoNi alloy, the magnetic effect is complicated by the presence of Mn atoms (due to strong magnetic frustrations of Mn). In other words, the CrMnFeCoNi alloy has a much stronger magnetism effect, as opposed to the NiCoCr alloy. Our atomistic model for NiCoCr yields similar behavior as the DFT calculations.

Experimentally, according to the report from an Oak Ridge group in Scientific Reports[31] in 2016, the antiferromagnetism of Cr "frustrates" the ferromagnetism of NiCo. The magnetic ordering does not show up at this composition all the way down to 2 K. The system behaves like a paramagnetic material with a susceptibility like Pd metal.

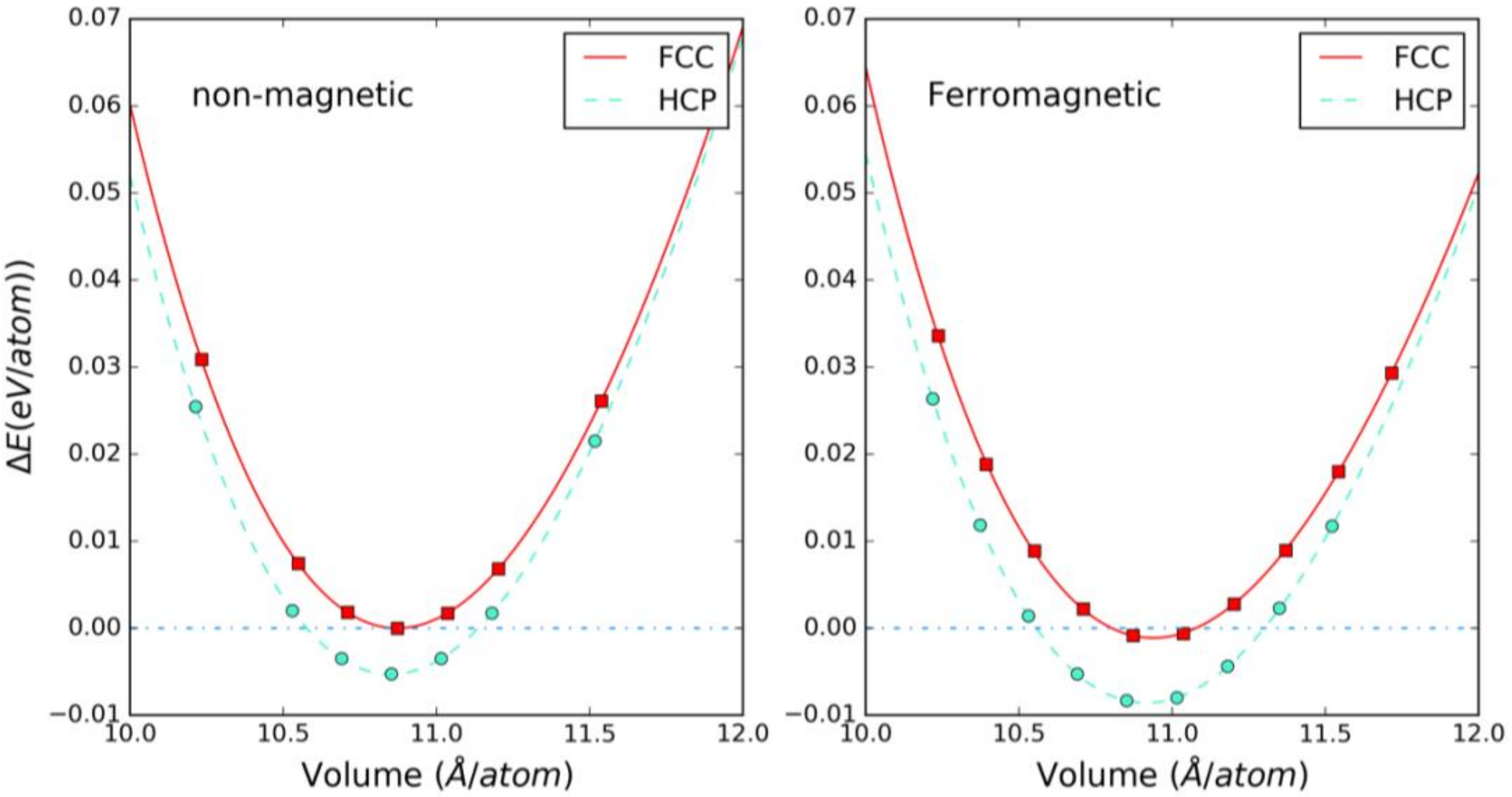

**Fig. S21 |** Results of ab initio calculations of the energy differences between fcc and hcp NiCoCr at 0K. Equation of states of fcc and hcp CrCoNi phases in the non-magnetic (left panel) and the ferromagnetic (right panel) states. The energies are plotted with reference to the non-magnetic fcc NiCoCr structure at the ground state. In both cases, the hcp structure is found to have a smaller formation energy. The magnetic contribution is found to be small in the NiCoCr alloy. The randomly populated super-cell approach was used to obtain the formation energies of fcc and hcp NiCoCr alloys, both in the magnetic state and in the non-magnetic state. The formation energies were averaged over 20 random configurations. For each calculation, the configuration (360 atoms) was geometrically optimized to reach the ground state at 0K.

But this 2016 experimental measurement also indicates that when the Cr content is lowered, magnetic ordering does become more obvious. When the alloy composition shifts to NiCoCr0.5 (Curie temperature rises to 250 K), our spin-polarized DFT calculations show that the energy difference between the magnetic and non-magnetic states becomes 15 meV/atom (and the volume difference is less than 2%). Therefore, even in this "obviously magnetic" case, the energy difference between the spin-polarized and non-polarized states is still relatively low, not up to the level that an EAM potential can resolve.

Magnetic effects may increase in our modeling when our solution alloy develops increasing local (partial) chemical order at low ageing temperatures. To evaluate the magnetic contribution in these configurations (each 360 atoms) with various LCOs (Figure S7), we performed spin-polarized DFT calculations on them. The configuration with the largest compositional variations shows an energy difference that is similar to that of the NiCoCr0.5 case above.

We mention here that the mechanical behavior of interest is mostly in the temperature range from room-temperature down to liquid nitrogen temperature. The zero-K potential energy landscape we describe is actually the finite-temperature free energy landscape with the thermal contribution subtracted from it. This energy landscape is not really the one at 0 K when some ferromagnetic state prevails, as finite temperature would randomize most of the spins.

All these said, we still have to admit that fundamentally, the EAM formalism is incapable of capturing complex magnetic effects at finite temperatures in a self-consistent manner. Some type of spin-dependent potential needs to be developed, but is beyond reach at present. As mentioned earlier, thus far EAM potentials have never been meant to monitor various degrees of magnetic ordering.

Therefore, the EAM potential as developed is an empirical NiCoCr-like atomistic model that enables a parametric study of the trend of dislocation behavior. This atomistic model is designed to capture the typical features of HEAs and MEAs: multi-principal (equiatomic) constituents, moderate chemical interactions, similar atomic sizes, single phase but with variable local chemical order, etc. The model is meant to analyze the trend due to LCO and consequences on dislocation responses, rather than pinning down energy numbers or nailing down the various contributions from chemical, elastic or magnetic energy terms. The model still overcomes a major hurdle in atomistic studies of high-entropy alloys: it allows large-scale (e.g., multi-million atoms simulations) MD modeling of dislocation activities, outside the realm of the more accurate first-principles DFT calculations.

**References for Supporting Information:**